\documentclass[a4paper,11pt]{article}
\usepackage{setspace}
\usepackage[top=2.54cm,bottom=2.54cm, left=2.54cm, right=2.54cm]{geometry}
\usepackage{bm}
\usepackage{natbib}
\usepackage{graphicx}
\usepackage{epstopdf}
\usepackage{amsmath}
\usepackage{hyperref}
\usepackage{caption}
\usepackage{amsfonts}

\begin{document}
\bibliographystyle{apa}

\begin{center}
\Large{A novel, divergence based, regression for compositional data}

\vspace{0.2cm}
\Large{M.~Tsagris \\
Independent researcher, Athens, \href{mailto:mtsagris@yahoo.gr}{mtsagris@yahoo.gr}}

\end{center}

\begin{center}
{\bf Abstract}
\end{center}
In compositional data, an observation is a vector with non-negative components which sum to a constant, typically 1. Data of this type arise in many areas, such as geology, archaeology, biology, economics and political science amongst others. The goal of this paper is to propose a new, divergence based, regression modelling technique for compositional data. To do so, a recently proved metric which is a special case of the Jensen-Shannon divergence is employed. A strong advantage of this new regression technique is that zeros are naturally handled. An example with real data and simulation studies are presented and are both compared with the log-ratio based regression suggested by Aitchison in 1986. \\
\\
\textbf{Keywords}: compositional data, Jensen-Shannon divergence, regression, zero values, $\phi$-divergence

\section{Introduction}
Compositional data is a special type of multivariate data in which the elements of each observation vector are non-negative and sum to a constant, usually taken to be unity. 
\begin{eqnarray*}
\mathbb{S}^d=\left\lbrace(x_1,...,x_D)|x_i \geq 0,\sum_{i=1}^Dx_i=1\right\rbrace  \ \ (d=D-1)
\end{eqnarray*}

A typical data set with $D=3$ components has the following form:
\begin{eqnarray*}
\left[ \begin{array}{ccc}
0.775   & 0.195   & 0.030 \\
0.719   & 0.249   & 0.320 \\
\vdots & \vdots & \vdots 
\end{array} \right]
\end{eqnarray*}

\begin{figure}[!h]
\includegraphics[scale=0.4,trim=0 60 0 50]{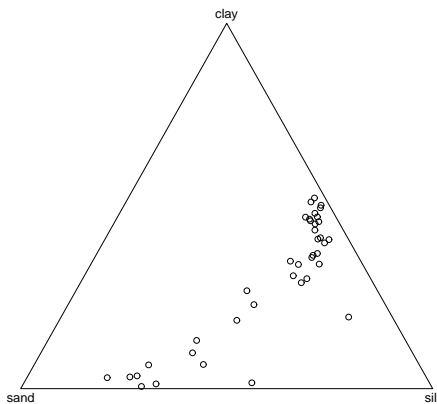}
\centering
\caption{Ternary diagram of the Arctic lake data (Aitchison, 2003).}
\label{figure}
\end{figure}

When $D=3$, a good way to visualize compositional data is the so called ternary diagram. Figure \ref{figure} presents the Arctic lake data (we will see this data set again later and describe it in more detail), which consist of three components, sand, silt and clay. These data re available in Aitchison (2003, p. 359) or the in R package \textit{compositions} along with more compositional data and techniques to analyze such data. Sand is a naturally occurring granular material composed of finely divided rock and mineral particles. Silt is a granular material of a size somewhere between sand and clay whose mineral origin is quartz and feldspar. As for clay, it is a fine-grained natural rock or soil material that combines one or more clay minerals with traces of metal oxides and organic matter. 

The closer a point is to a vertex, the higher its value in that corresponding component. If the point is on the vertex, it means that its value in the component corresponding to that vertex is 1 and the other two values are 0. On the opposite side, the further away a point is from a vertex, the lowest its value in the corresponding component. If a point lies on an edge, it means, that the value of the value of the component corresponding to that vertex is zero. If the point is on the middle of that edge, the values of the other two components are exactly 0.5. Points at the barycenter indicate that the values are  equal (to 1/3). For example, the points lying at the base of the triangle in Figure \ref{figure} have very low values of clay and higher values sand than silt. 

Analysis of such data may be carried out in various ways which can be summarized in two directions: either by transforming the data or not. In the first direction, the most popular transformations are the log of ratios formed by the components (Aitchison, 1982; 2003). Other approaches include taking the square root of the data, resulting in data which lie on the hypersphere (Stephens, 1982 and  Scealy \& Welsh, 2011). Another parametric model of this school of thought is the Dirichlet distribution (Gueorguieva et al., 2008). In all of these cases, regression models have been developed.      

An important issue in compositional data is the presence of zeros, which cause problems for the logarithmic transformation. The issue of zero values in some components is not addressed in most papers, but see Neocleous et al.(2011) for an example of discrimination in the presence of zeros. Only when treated as directional data, compositional data have no problem with zeros. Log-ratio and Dirichlet regression zero imputation techniques must be applied prior to fitting the models. 

The paper proposes a novel regression method based on a recently suggested metric. It is a new metric for probability distributions (Endres \&  Schindelin, 2003 and {\"O}sterreicher \& Vajda, 2003), which is a special case of the Jensen-Shannon divergence. We will use it for compositional data, since they also sum to $1$. The main advantage of this newly proposed regression is that it handles zeros naturally and in addition it can lead to better fits as will be seen later. 

The paper is structured as follows. Section 2 reviews some regression approaches for compositional data analysis and introduces the new regression model. Section 3 contains an example using real data and simulation studies to illustrate the performance of the newly proposed regression. Some Miscellanea and the Conclusion appear in Section 4 and 5 respectively.  

\section{Regression models}
The most popular approach is to use the logistic normal distribution (Aitchison, 1982; 2003)
\begin{eqnarray*} 
\log\left(\frac{{\bf y}_{-D}}{y_D}\right)=\left(\log{\frac{y_1}{y_D}},\ldots,\log{\frac{y_d}{y_D}}\right) \sim  \text{N}_d\left({\bf x}^T{\bf B},\pmb{\Sigma}\right), 
\end{eqnarray*}
where ${\bf x}^T$ is a column vector of the design matrix ${\bf X}$, $D$ is the number of components, $d=D-1$ and 
\begin{eqnarray*}
{\bf B}=\left(\pmb{\beta}_1,\ldots,\pmb{\beta}_d\right) \ \ \text{and} \ \ \pmb{\beta}_i=\left(\beta_{0i},\beta_{1i},...,\beta_{pi} \right)^T, \ i=1,...,d.
\end{eqnarray*}
are the regression coefficients and $p$ is the number of independent variables. We will denote this regression as Aitchison regression. 

A Dirichlet distribution (Gueorguieva et al., 2008) with the parameters linked to some covariates is another approach
\begin{eqnarray*}
{\bf y} \sim  \text{Dir}\left(\frac{\phi}{1+\sum_{j=2}^De^{{\bf x}^T\pmb{\beta}_j}}, \frac{\phi e^{{\bf x}^T\pmb{\beta}_2}}{1+\sum_{j=2}^De^{{\bf x}^T\pmb{\beta}_j}}, \ldots, \frac{\phi e^{{\bf x}^T\pmb{\beta}_d}}{1+\sum_{j=2}^De^{{\bf x}^T\pmb{\beta}_j}} \right). 
\end{eqnarray*}
Note, that the precision parameter $\phi$ can also be linked to the independent variables via link function of course to ensure positivity of the estimated values. Maier (2014) has written an R package to fit Dirichlet regression models with and without covariates on $\phi$. 

Two approaches were examined in Murteira \& Ramalho (2014) are the OLS regression
\begin{eqnarray*}
\min_{\pmb{\beta}} \sum_{i=1}^n\left[{\bf y}_i-{\bf f}_i\left(\pmb{\beta};{\bf x}\right)\right]\left[{\bf y}_i-{\bf f}_i\left(\pmb{\beta};{\bf x}\right)\right]
\end{eqnarray*}
and the multinomial logit regression
\begin{eqnarray*}
\min_{\pmb{\beta}} \sum_{i=1}^ny_i\log{\frac{y_i}{{\bf f}_i\left(\pmb{\beta};{\bf x}\right)}}=
\max{\pmb{\beta}} \sum_{i=1}^ny_i\log{{\bf f}_i\left(\pmb{\beta};{\bf x}\right)}, 
\end{eqnarray*}
where
\begin{eqnarray*}  
{\bf f}_i\left(\pmb{\beta};{\bf x}\right)=\left(\frac{1}{\sum_{j=1}^De^{{\bf x}_i^T\pmb{\beta}_j}},
\frac{e^{{\bf x}_i^T\pmb{\beta}_2}}{\sum_{j=1}^De^{{\bf x}_i^T\pmb{\beta}_j}},\ldots, 
\frac{e^{{\bf x}_i^T\pmb{\beta}_d}}{\sum_{j=1}^De^{{\bf x}_i^T\pmb{\beta}_j}}\right).
\end{eqnarray*}

Finally, the Kent distribution Kent (1982) was employed by Scealy \& Welsh (2011) after the square root was applied component-wise to the data, thus ampping them to the unit (hyper)-sphere
\begin{eqnarray*} 
\sqrt{\bf y} \sim \text{Kent}\left(\pmb{\mu}\left(\pmb{\beta},{\bf x}\right),\pmb{\Gamma},\kappa,\pmb{\delta}\right).
\end{eqnarray*}

\subsection{\hspace{-0.3cm}The ES-OV regression}
We advocate that as a measure of the distance between two compositions we can use a special case of the Jensen-Shannon divergence
\begin{eqnarray} \label{esov} 
\text{ES-OV}\left({\bf x},{\bf y}\right)=\sum_{j=1}^D\left( x_j\log{\frac{2x_j}{x_j+y_j}}+y_j\log{\frac{2y_j}{x_j+y_j}} \right),
\end{eqnarray}
where ${\bf x}$ and  ${\bf y} \in \mathbb{S}^d$. Endres \& Schindelin (2003) and {\"O}sterreicher \& Vajda (2003) proved, independently, that (\ref{esov}) satisfies the triangular identity and thus it is a metric. The names ES-OV comes from the researchers' initials. 
In fact, (\ref{esov}) is the square of the metric, still a metric, and we will use this version.

The idea is simple and straightforward, minimization of the ES-OV metric between the observed and the fitted compositions with respect to the beta coefficients
\begin{eqnarray} \label{esovreg} 
\min_{\pmb{\beta}} \sum_{i=1}^D\left( {\bf y}_i\log{\frac{2{\bf y}_i}{{\bf y}_i+{\bf f}_i\left(\pmb{\beta};{\bf x}\right)}}+
{\bf f}_i\left(\pmb{\beta};{\bf x}\right)\log{\frac{{\bf f}_i\left(\pmb{\beta};{\bf x}\right)}{{\bf y}_i+{\bf f}_i\left(\pmb{\beta};{\bf x}\right)}} \right).
\end{eqnarray}

Below we summarise a few properties of the ES-OV metric and its associated regression model. \\
\begin{itemize}
\item The ES-OV metric belongs to the class of $\phi$-divergences. Let $f\left(t\right)=t\log{\frac{2t}{1+t}}+\log{\frac{2}{1+t}}$, then 
$\text{ES-OV}\left({\bf y},{\bf z}\right)=\sum_{j=1}^Dz_j f\left(\frac{y_j}{z_j}\right)$. Therefore, we can say that this regression falls within the minimum $\phi$-divergence regression algorithms. 
\item A weighted version of the ES-OV regression (\ref{esovreg}), such as \\
$\sum_{i=1}^D\left( \lambda {\bf y}_i\log{\frac{2{\bf y}_i}{{\bf y}_i+\hat{{\bf y}}_i}}+\left(1-\lambda\right) \hat{{\bf y}}_i\log{\frac{2\hat{{\bf y}}_i}{{\bf y}_i+\hat{{\bf y}}_i}} \right)$ (Jensen-Shannon divergence) produces the best fits when $\lambda=0$.
\item A similar version of (\ref{esovreg}) is $\sum_{i=1}^D\left( y_i\log{\frac{y_i}{\hat{y}_i}}+\hat{y}_i\log{\frac{\hat{y}_i}{y_i}} \right)$, which gives nice results, but not as good (\ref{esovreg}).
\item The ES-OV regression (\ref{esovreg}) can deal with zero values naturally as already mentioned in the Introduction. 
\item The ES-OV regression (\ref{esovreg}) can lead to better fits than the logistic normal. \\
\end{itemize}

The consistency and the asymptotic distribution of the regression parameters has not been studied. This is a task we have to do and we have confidence that the answers will be positive. As for the last property and in general the fit of regression models for compositional data we suggest the use of the Kulback-Leibler divergence (Kullback, 1997), which was also used as a measure of fit by Theil (1967)
\begin{eqnarray} \label{kl}
\sum_{i=1}^n\text{KL}\left({\bf y}_i,\hat{{\bf y}}_i\right)=\sum_{i=1}^n{\bf y}_i\log{\frac{{\bf y}_i}{\hat{\bf y}_i}}. 
\end{eqnarray} 

\section{Example with real data and simulation studies}
In both the example and the simulation studies we will compare the ES-OV regression with the Aitchison regression. 

\subsection{Real data analysis}
\begin{figure}[!ht]
\centering
\begin{tabular}{cc} 
\includegraphics[scale=0.4,trim=0 80 0 50]{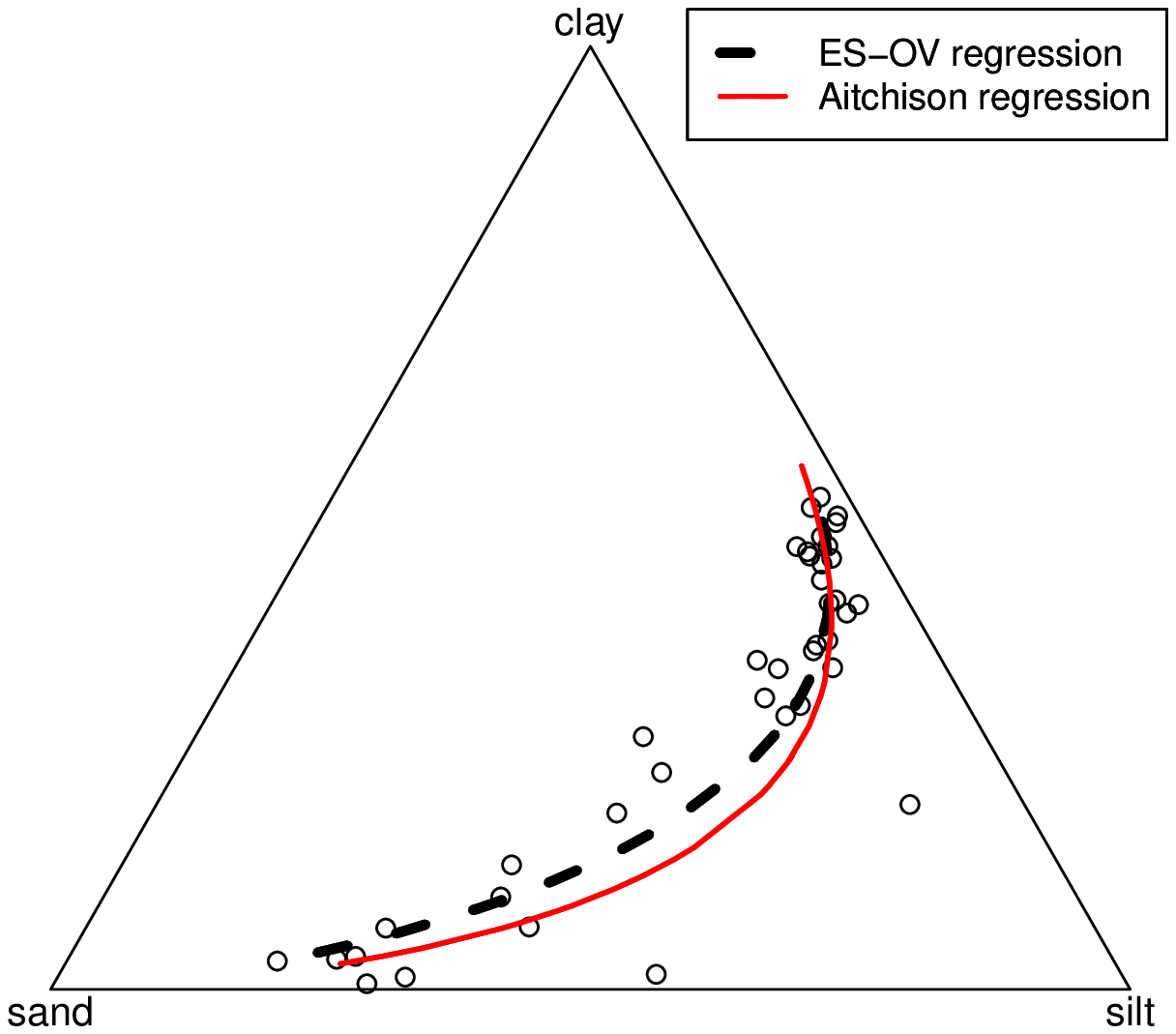} &
\includegraphics[scale=0.4,trim=0 80 0 50]{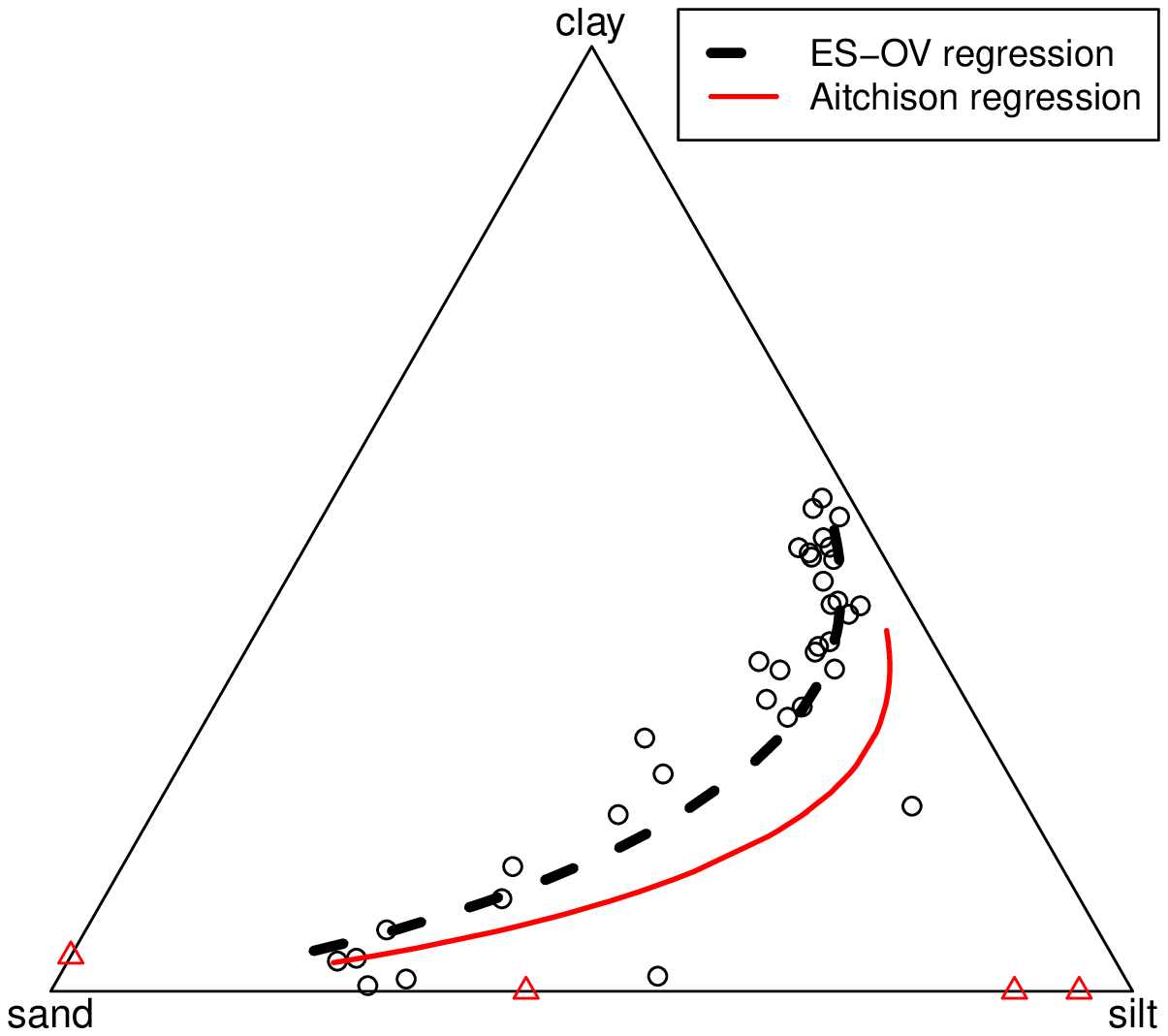} \\
(a)   & (b)   \\
\end{tabular}
\caption{Regression lines for the Arctic lake data. The ternary diagram on the left contains the data as they are, whereas the right diagram has some zero values added.}
\label{reg1}
\end{figure}

\begin{figure}[!ht]
\centering
\begin{tabular}{ccc}
\multicolumn{3}{c}{Arctic lake data} \\ 
\includegraphics[scale=0.35,trim=0 20 0 20]{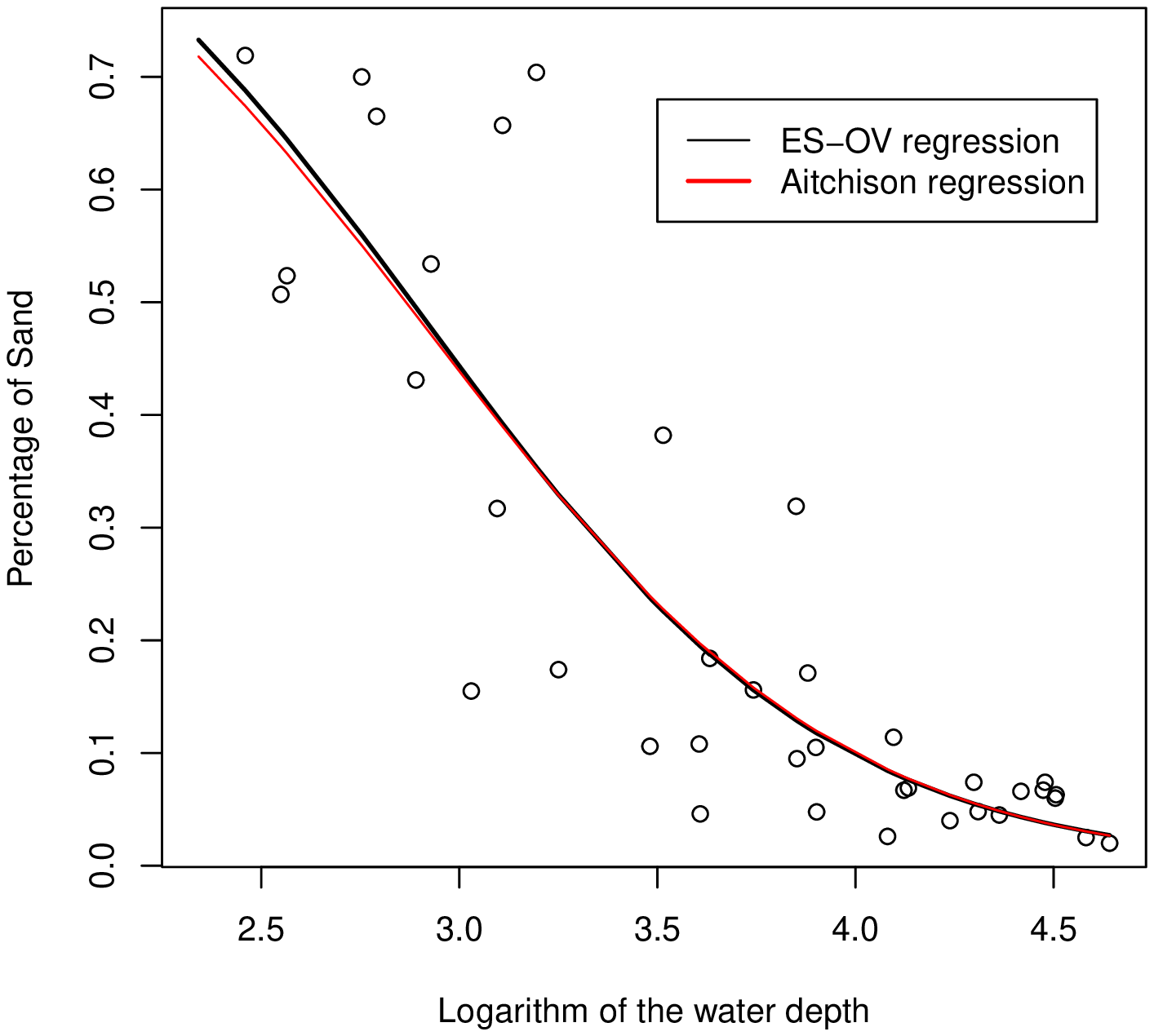} &
\includegraphics[scale=0.35,trim=0 20 0 20]{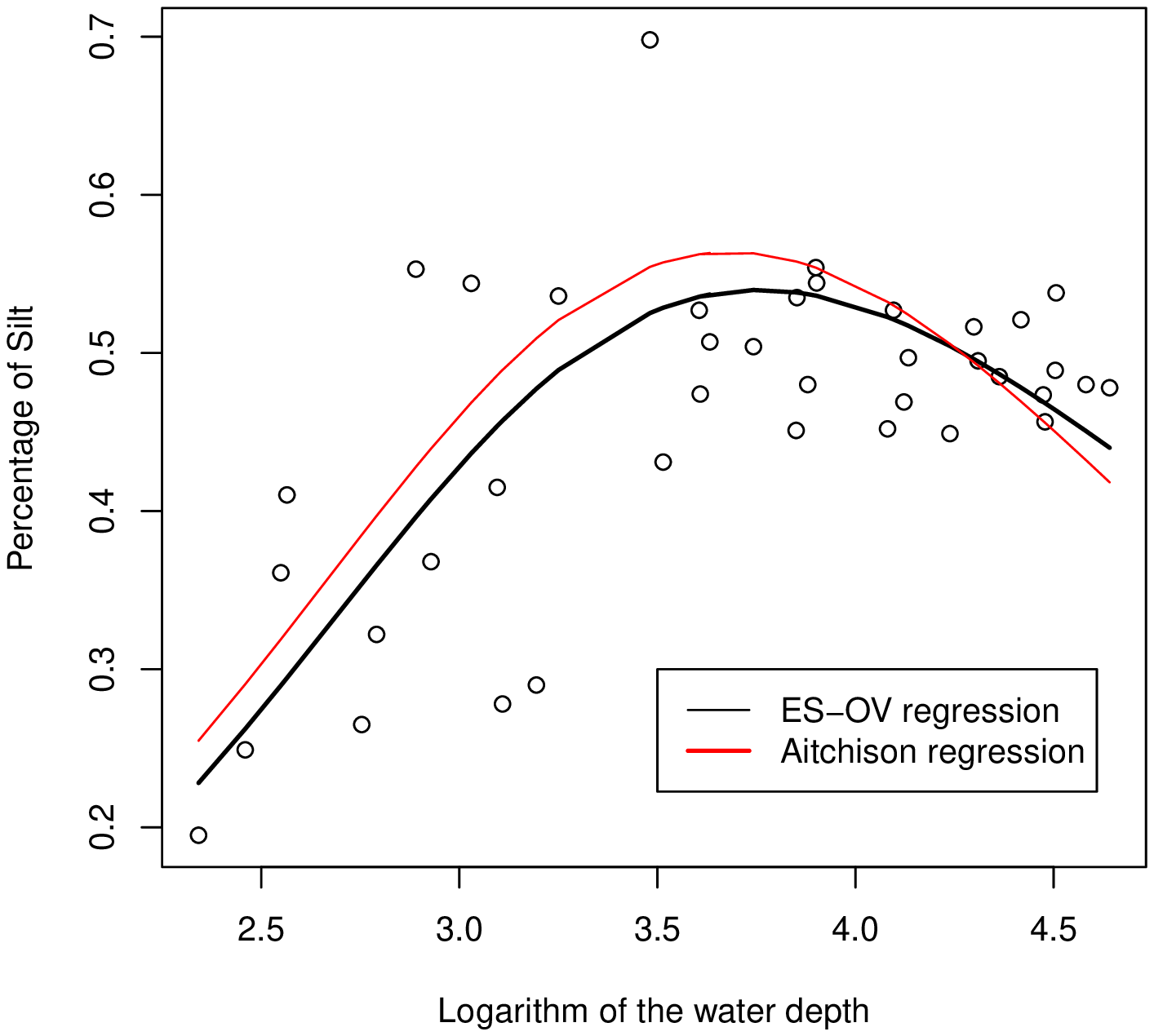} &
\includegraphics[scale=0.35,trim=0 20 0 20]{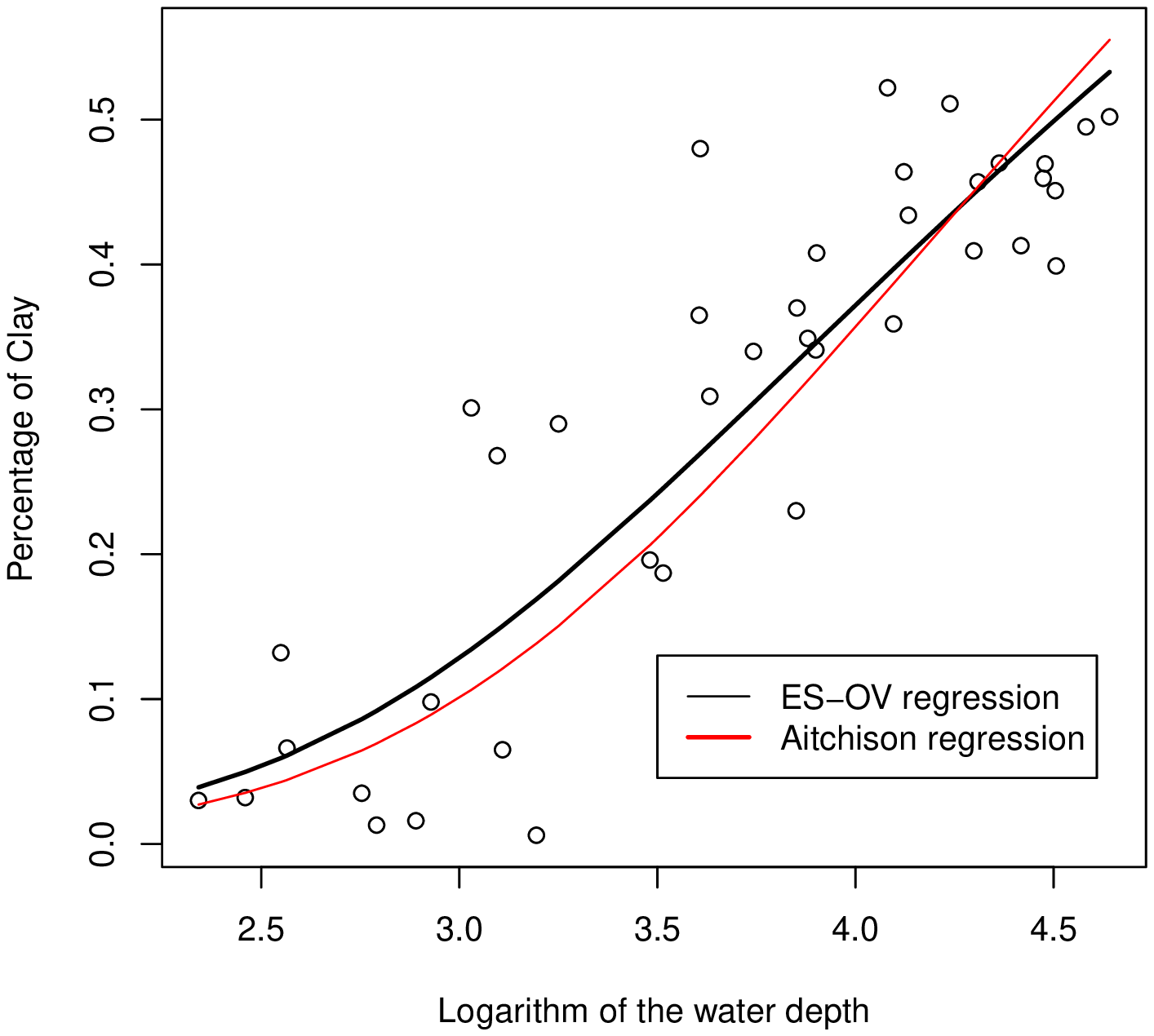} \\
Sand    &  Silt   &   Clay  \\
        &         &         \\
\multicolumn{3}{c}{Arctic lake data with zeros} \\ 
\includegraphics[scale=0.35,trim=0 20 0 20]{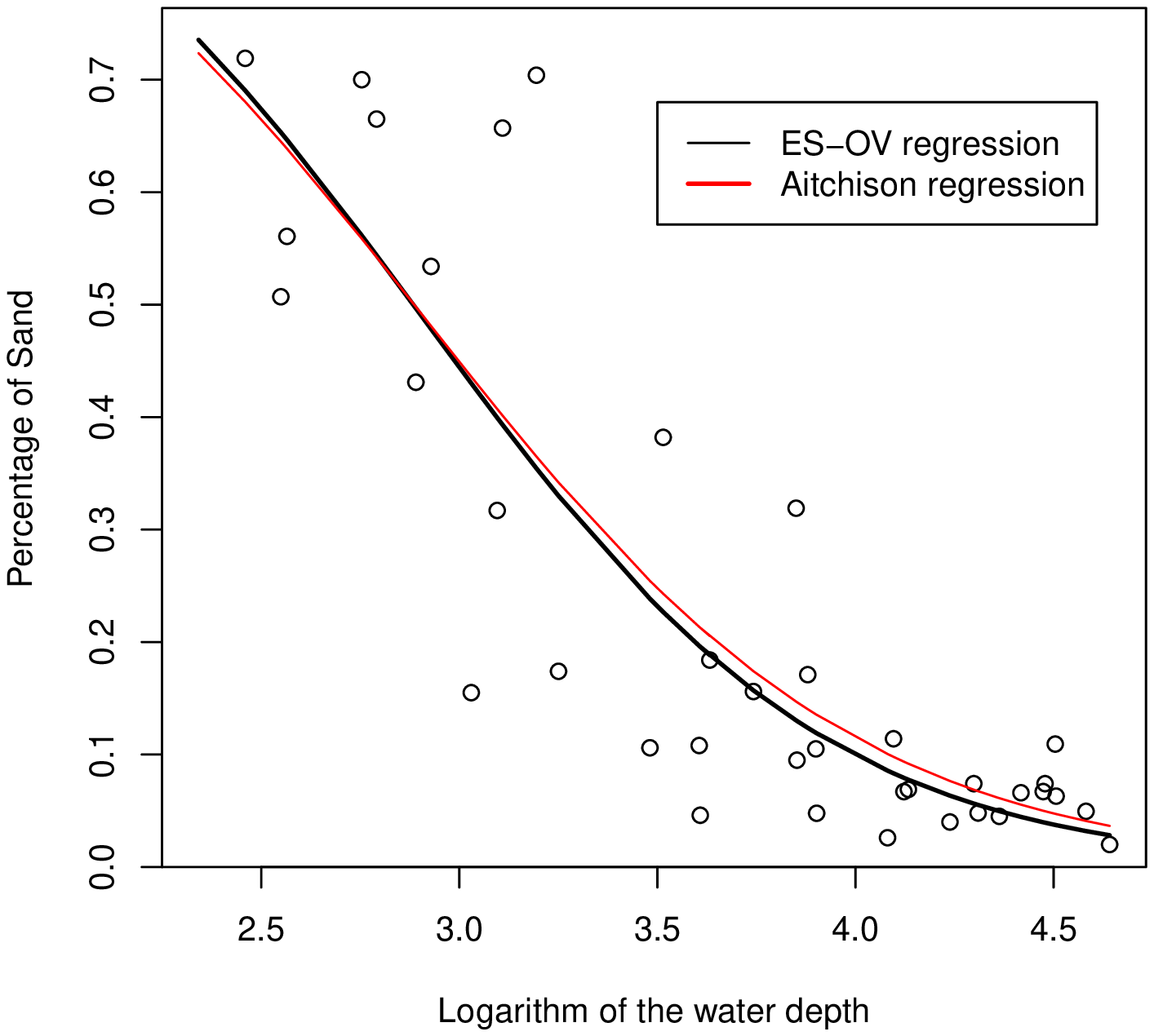} &
\includegraphics[scale=0.35,trim=0 20 0 20]{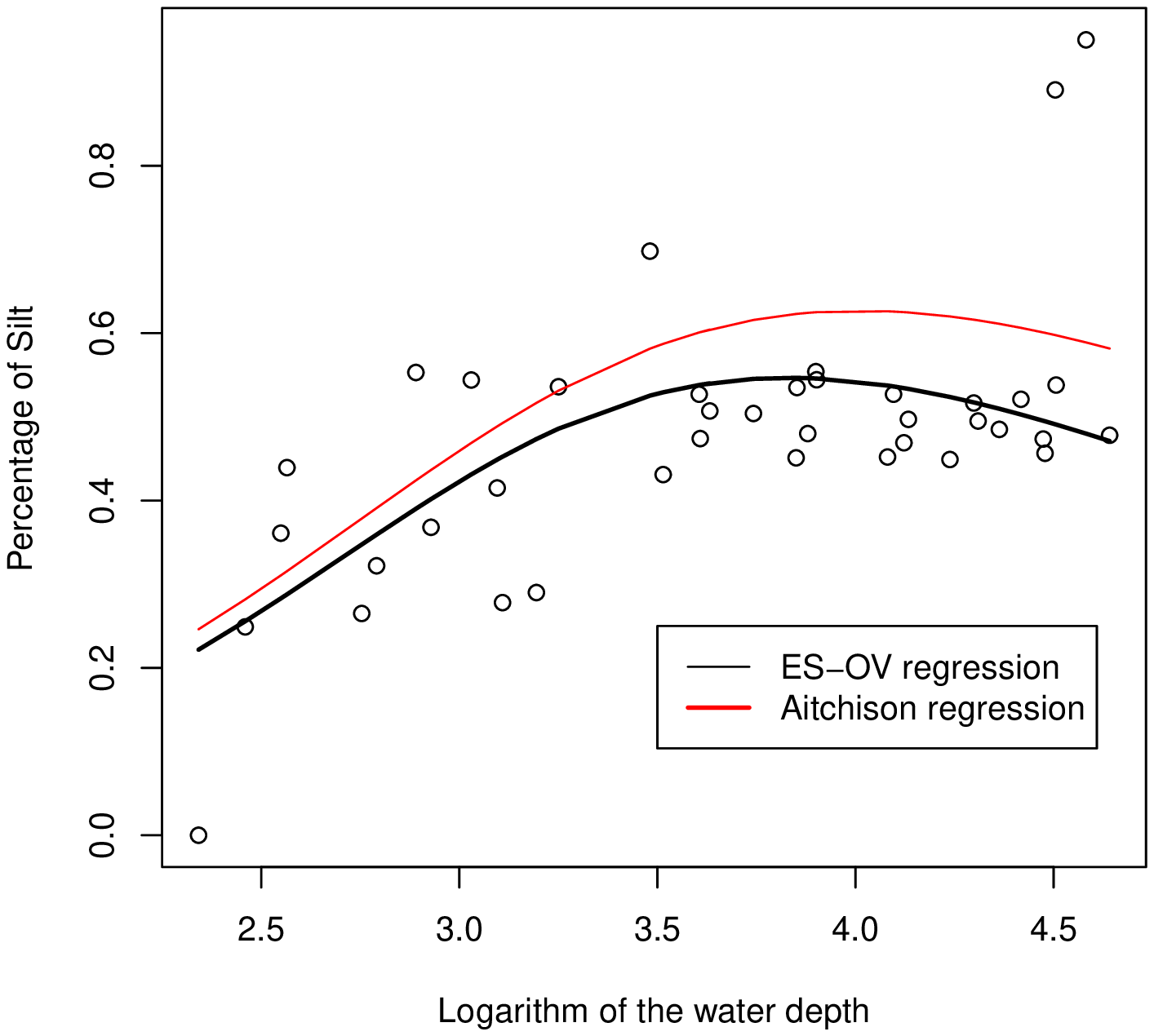}  &
\includegraphics[scale=0.35,trim=0 20 0 20]{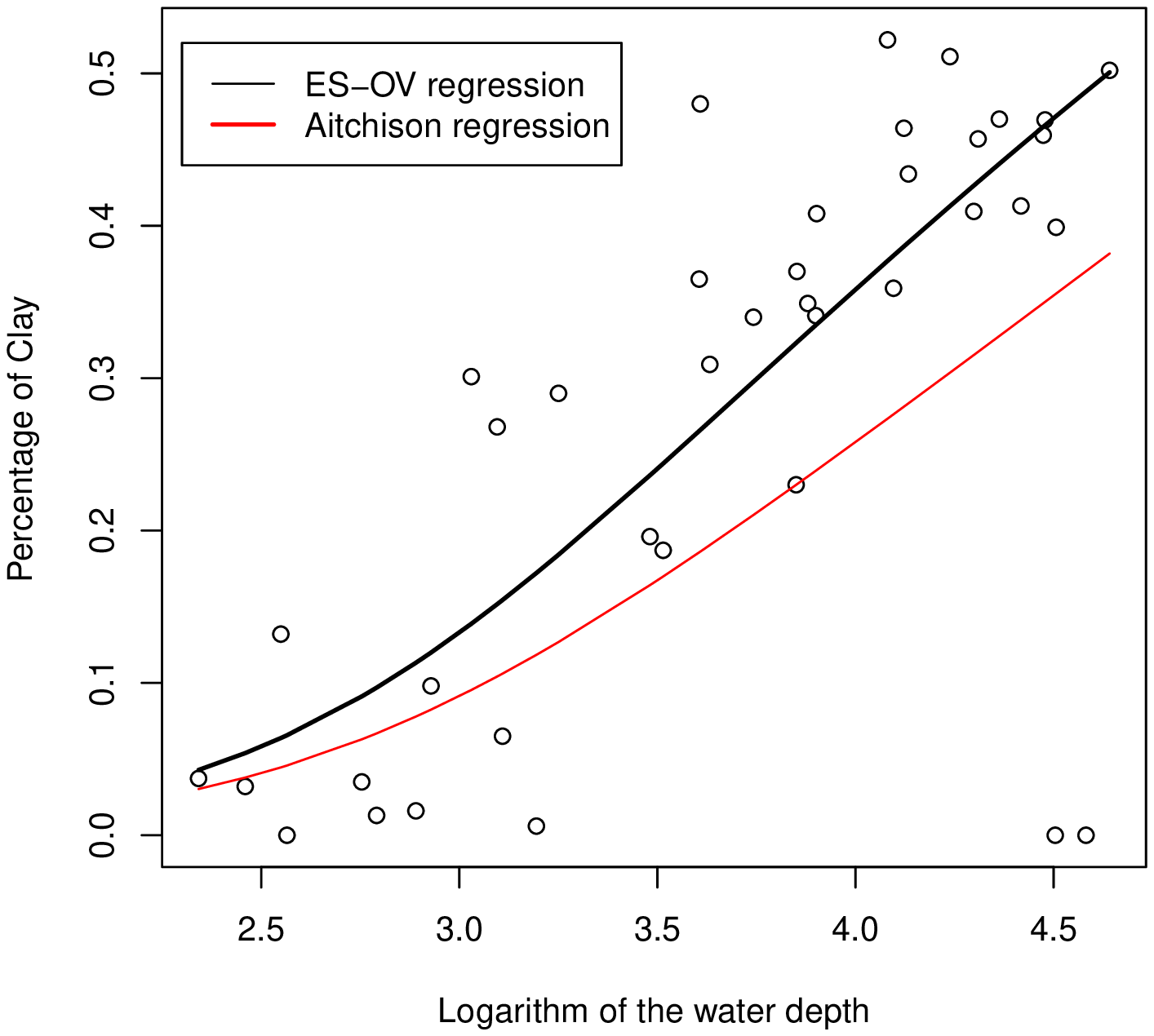} \\
Sand    &  Silt &   Clay  \\
\end{tabular}
\caption{All graphs contain the logarithm of the water depth against the observed and the estimated percentages of the three components. The graphs in first row correspond to the observed Arctic lake data, while the graphs in the second row correspond to the data with some zeros added.}
\label{reg1a}
\end{figure}

In the first instance we will present an example with the Arctic lake (real) data we saw in Figure 1. Three ingredients, \textit{sand}, \textit{silt} and \textit{clay} were measured at various depths (39 measurements) of an Arctic lake (Aitchison, 1986, p.359). The goal is to see how the composition in these three elements changes as the water depth increases and how good can our predictions be. We can see in Figure \ref{reg1}(a) that both methods fit the data well. As we move from left to right, we go from the surface to deeper levels of the lake. The composition of the samples has initially high percentage of sand, but as we move to the bottom of the lake, this percentage reduces, while the percentages of silt and clay increase. This makes absolute sense, since the percentages sum to $1$, the derivatives, or the rates of change of the components must sum to zero. Figure \ref{reg1a} serves as an ancillary to Figure \ref{reg1} for the readers who are not familiar with the ternary diagrams. 

We then make the values of four components equal to zero. In specific, the percentage of clay becomes zero for three observations and the percentage of silt becomes zero for one observation. In order to apply the logistic normal regression, imputation or zero replacement techniques (Mart{\'\i}n-Fern{\'a}ndez, 2012) are required. Templ et al. (2011) has created an R package (\href{http://cran.r-project.org/web/packages/robCompositions/index.html}{robCompositions}) which performs zero value replacement for compositional data. Figure (\ref{reg1}(b)) shows the Arctic lake data with the zero values (red triangles in the edges of the triangle). The Aitchison regression, applied to the new, zero value replaced data has been heavily affected by the zero values and the fitted values fall outside the bulk of the data. The ES-OV regression on the other hand seems unaffected. 

\subsection{Simulation studies}
Simulation studies were conducted in the spirit of prediction performance. As we mentioned before, the asymptotic properties of the ES-OV regression, thus we cannot use it for inference about the parameters. For this reason, we can only use it for prediction purposes. This is the current drawback of this regression. 

We generated data from logistic normal regression models with various sample sizes  ($n=25,50,75,100$) and components ($d=6,11,16$). We repeated this scenario and made about half of the observations (simulating from a uniform variable) each time contain zero values. There were $2$, $4$ and $6$ components with zero values when the simulated data consisted of $6$, $11$ and $16$ components. For each case, we applied one-fold cross validation. That is, we remove one observation and estimate the regression parameters using the remaining data. Then we predict the value of the observation and calculate the Kullback-Leibler divergence of the true from the fitted value. This was repeated for all observations and in the end we summed all the divergences
\begin{eqnarray*}
\text{KL}\left({\bf y},\hat{\bf y}\right)=\sum_{i=1}^n{\bf y}_i\log{\frac{{\bf y}_i}{\hat{\bf y}_{-i}}},
\end{eqnarray*}
where $\hat{\bf y}_{-i}$ is the $i$-th predicted compositional observation with the $i$-th observation having been excluded from the estimation of the regression parameters.

The one-fold cross validation procedure was repeated $200$ iterations, due to limited computational sources, for each combination of sample size and number of components. In the end we averaged the Kulback-Leibler divergences
\begin{eqnarray} \label{akl}
D=\frac{1}{200}\sum_{j=1}^{200}\text{KL}_j\left({\bf y},\hat{\bf y}\right). 
\end{eqnarray}

The number $200$ might seem small, but I think is large enough to make valid conclusions. The sample sizes considered are small, again due to limited computational resources. Even in this case, we believe we can extract some valid conclusions.  

\subsubsection*{Compositional data with no zeros}
The pseudo-code for the first set of the simulation studies is given below
\begin{enumerate}
\item[Step 1.] Generate $n$ data from a multivariate normal regression model ${\bf Z} \sim N_p\left({\bf BX},\pmb{\Sigma}\right)$,
where ${\bf X}$ is a design matrix with 2 independent variables. $\bf B$ were chosen randomly from a standard normal distribution and $\pmb{\Sigma}$ was a diagonal matrix with variances generated from an $\text{Exp}(1)$ distribution.
\item[Step 2.] Make the ${\bf Z}$ compositional using
\begin{eqnarray*}
y_1 = \frac{1}{1+\sum_{j=1}^pe^{z_j}}, \ \ y_i = \frac{e^{z_i}}{1+\sum_{j=1}^pe^{z_j}}, \ \ \text{for} \ \ i=2,\ldots,p.
\end{eqnarray*}
\item[Step 3.] Perform the ES-OV and Aitchison regressions and calculate (\ref{akl}) each time. 
\item[Step 4.] Repeat Steps 1-3 for various small sample sizes $n=(25,50,75,100)$ and number of variables $p=(5,10,15)$. Thus, we now have $D=(6,11,16)$ number of components and essentially $15, 30$ and $45$ beta parameters to estimate. 
\end{enumerate}

\begin{figure}[!ht]
\centering
\begin{tabular}{cccc}
\multicolumn{4}{c}{6 components} \\ 
\includegraphics[scale=0.25,trim=0 20 0 20]{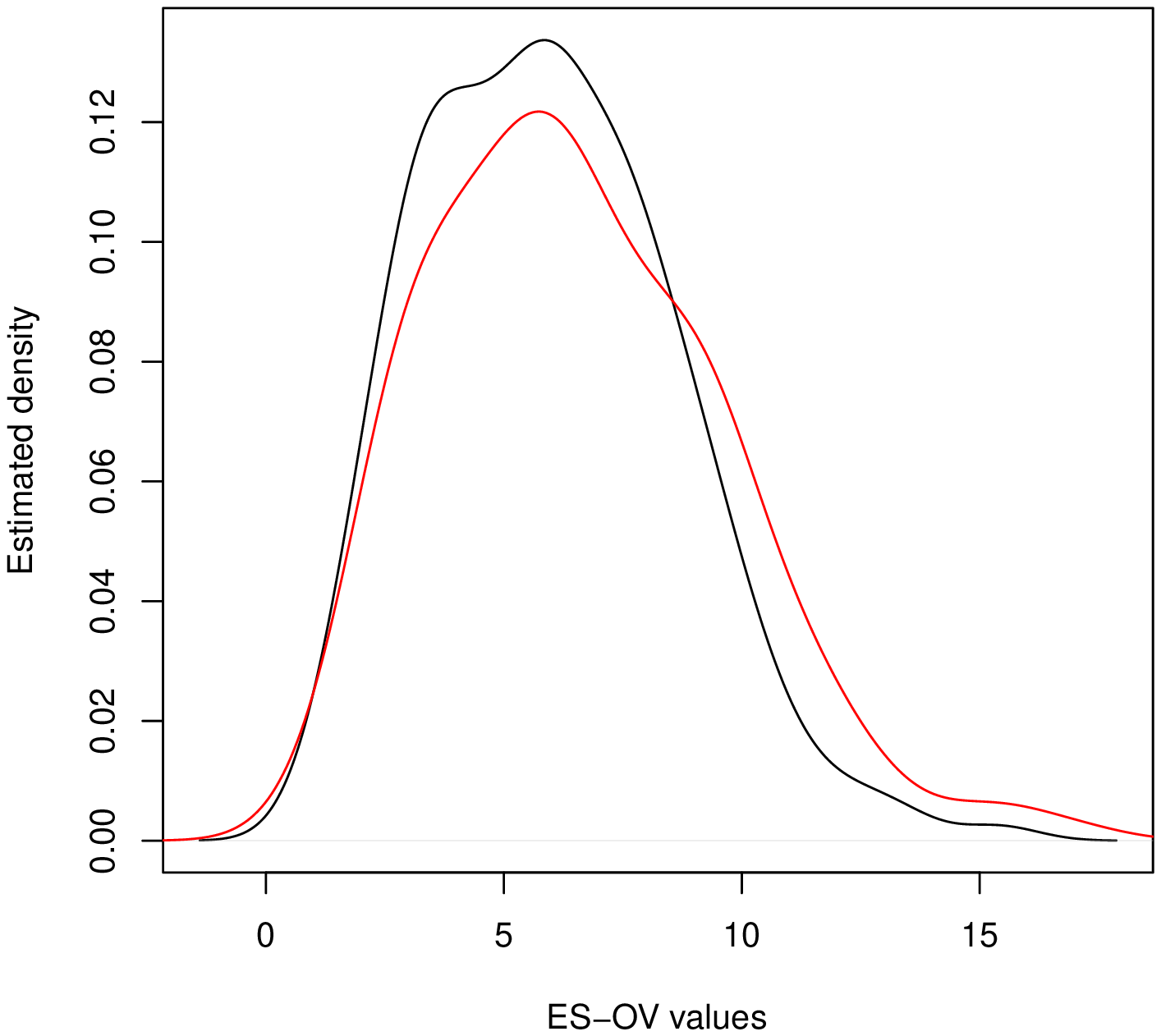} &
\includegraphics[scale=0.25,trim=0 20 0 20]{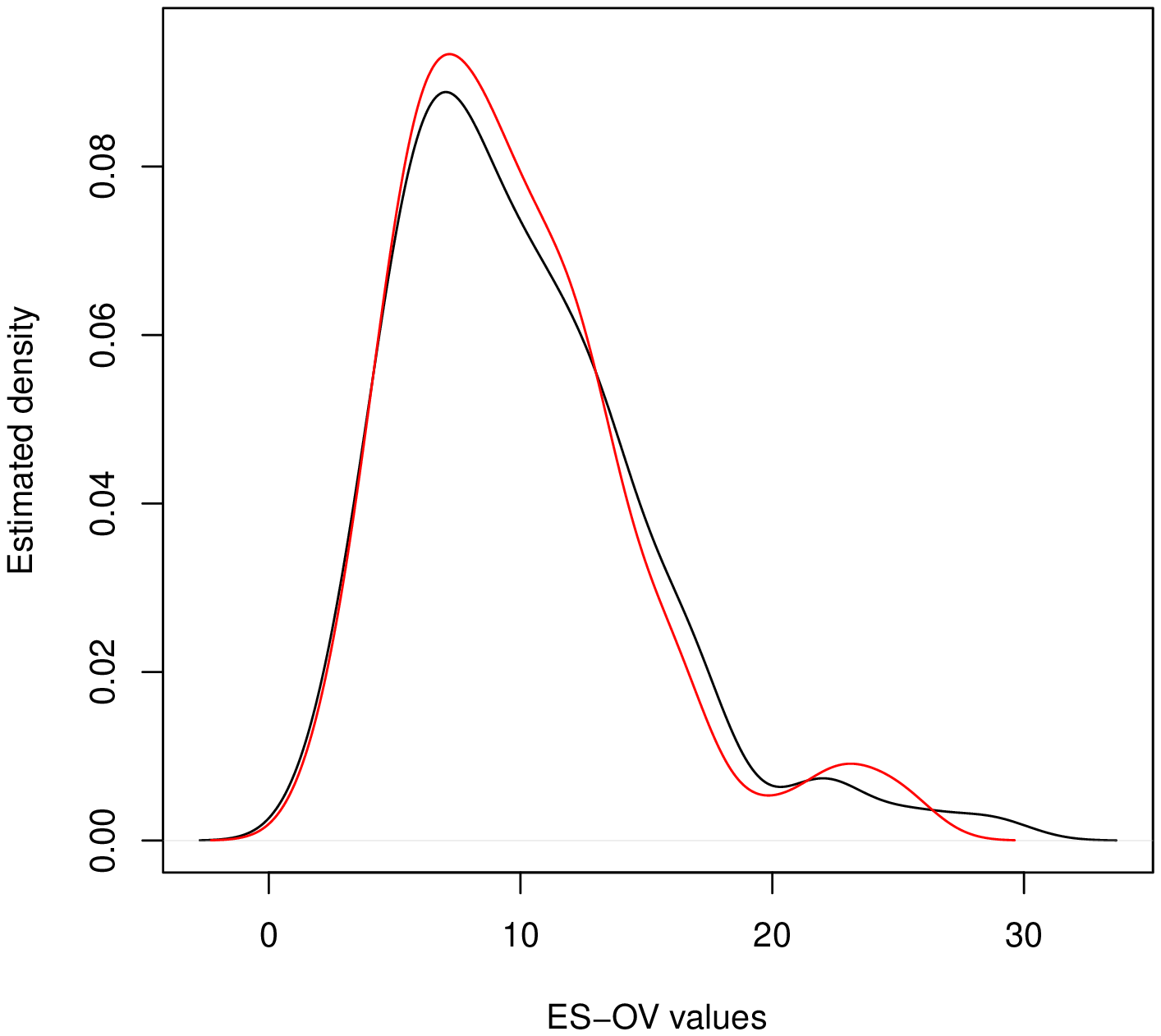} &
\includegraphics[scale=0.25,trim=0 20 0 20]{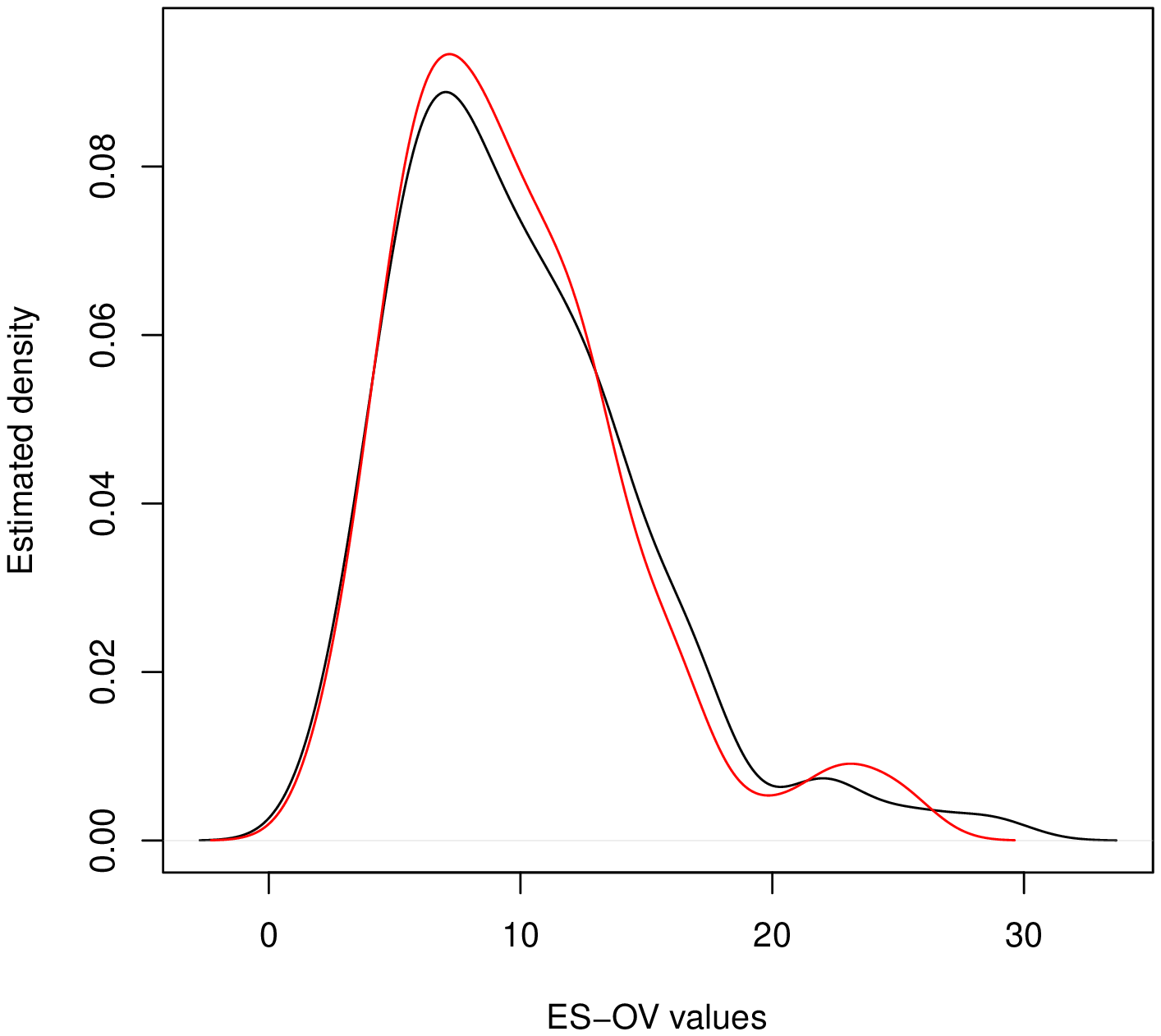} &
\includegraphics[scale=0.25,trim=0 20 0 20]{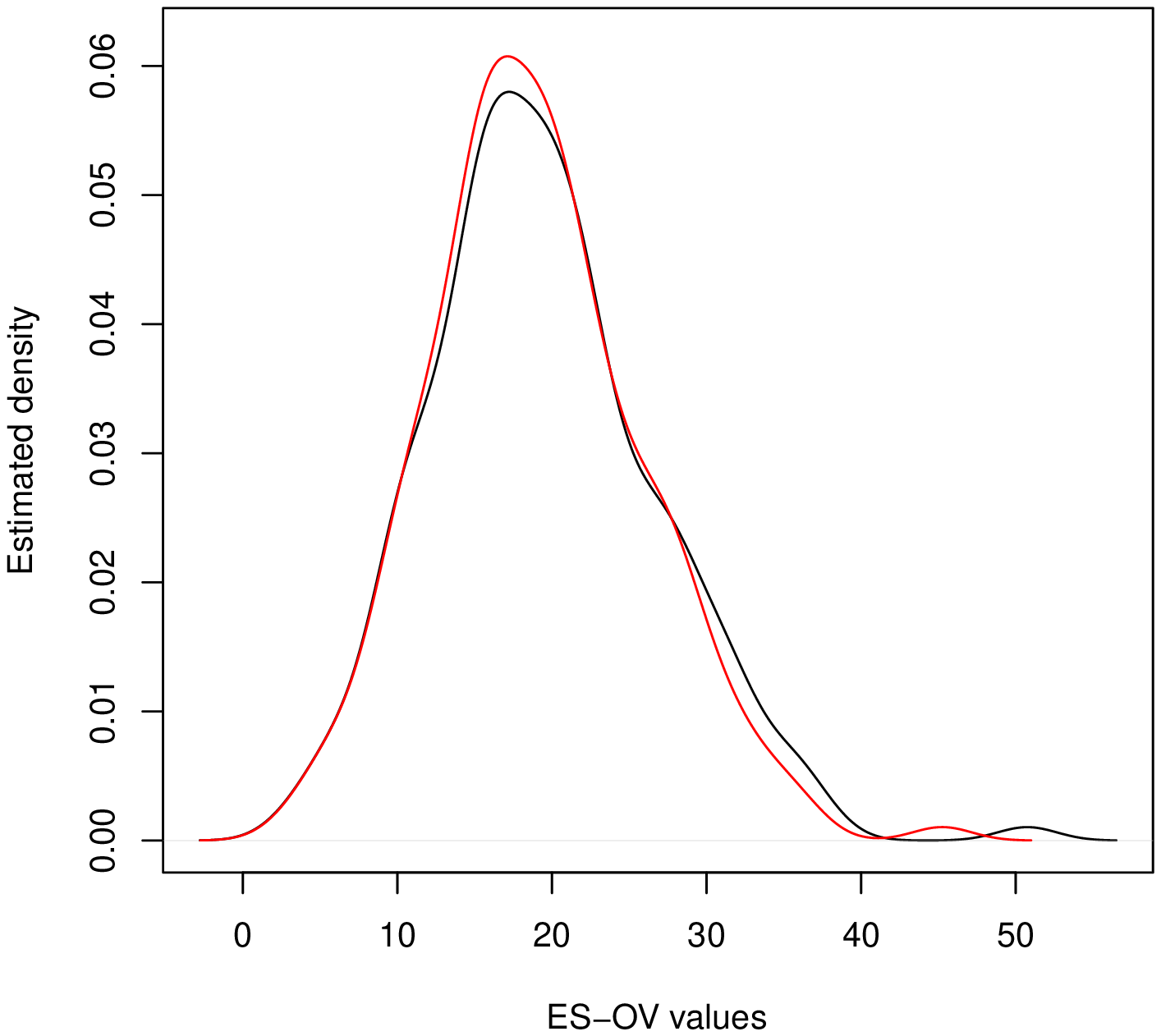} \\
$n=25$ & $n=50$ & $n=75$ & $n=100$                     \\
\multicolumn{4}{c}{11 components} \\ 
\includegraphics[scale=0.25,trim=0 20 0 20]{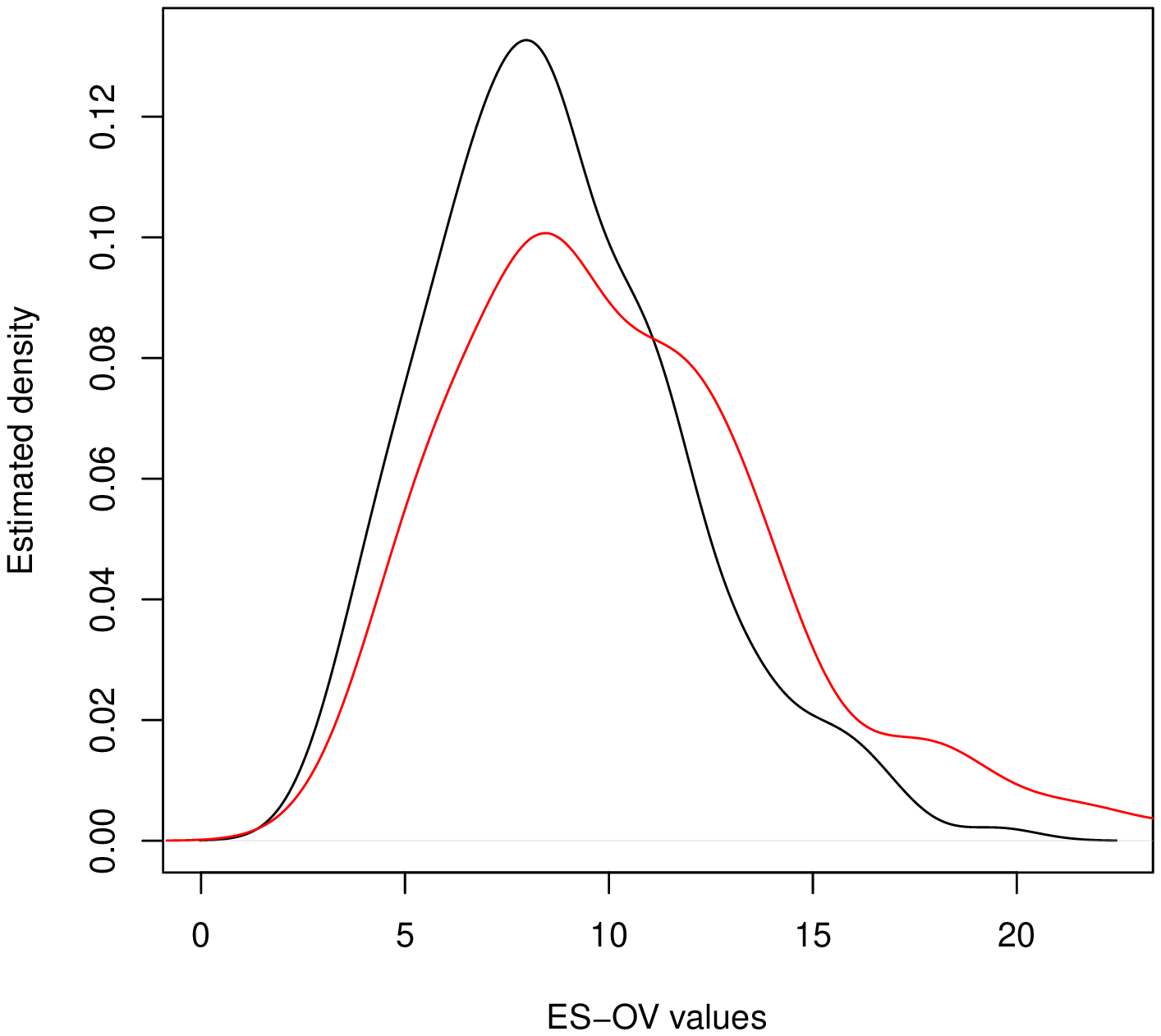} &
\includegraphics[scale=0.25,trim=0 20 0 20]{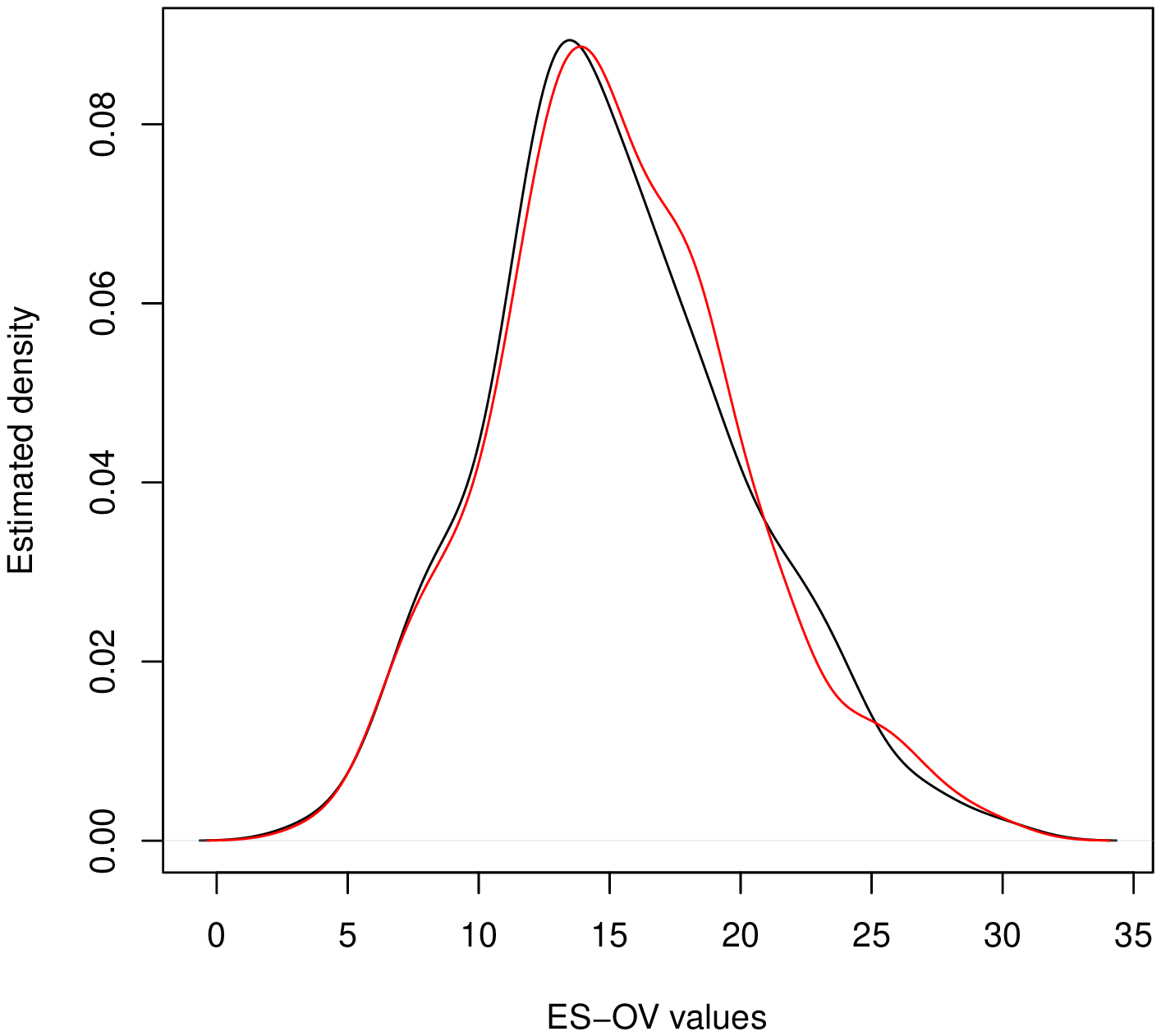} &
\includegraphics[scale=0.25,trim=0 20 0 20]{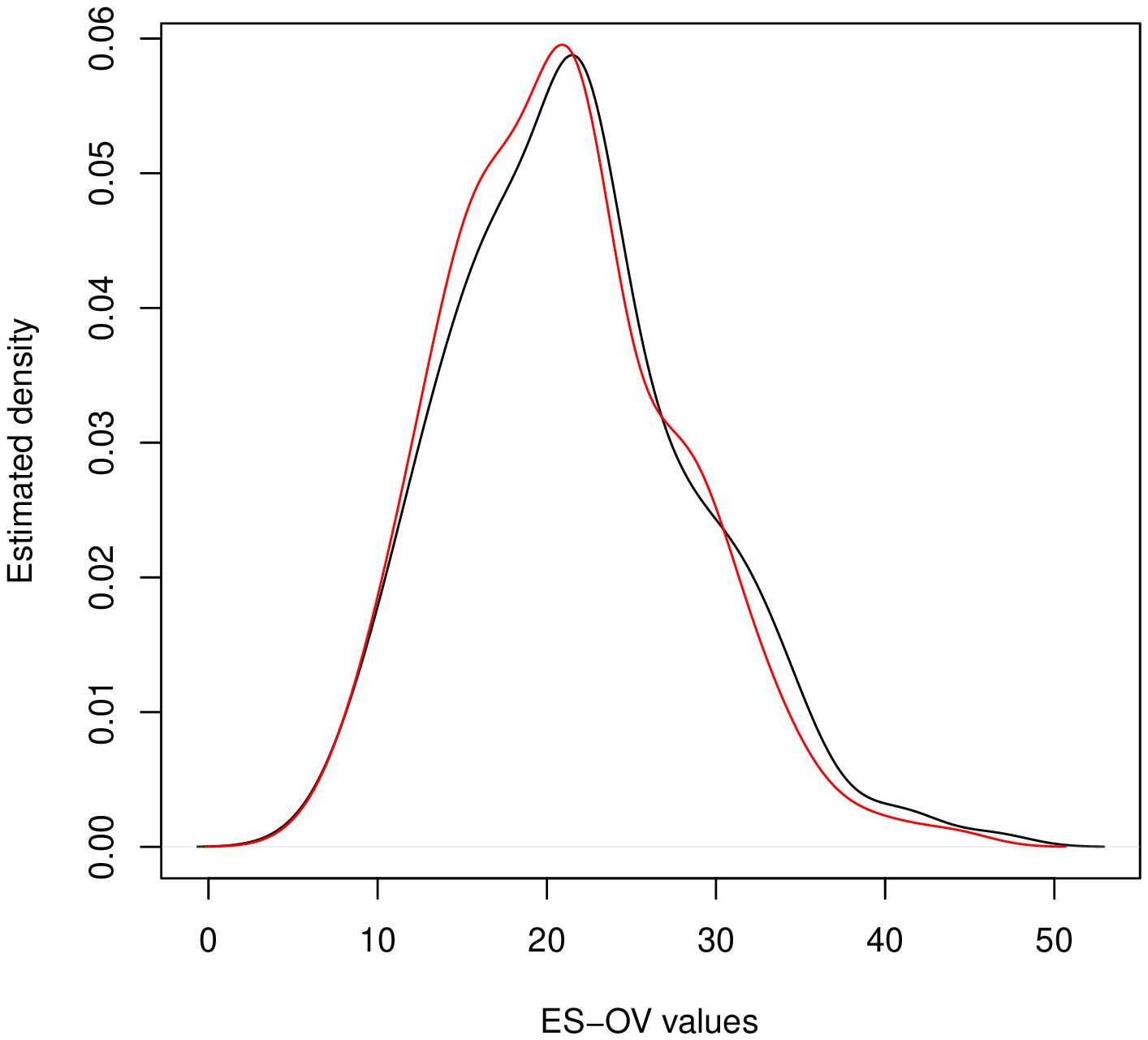} &
\includegraphics[scale=0.25,trim=0 20 0 20]{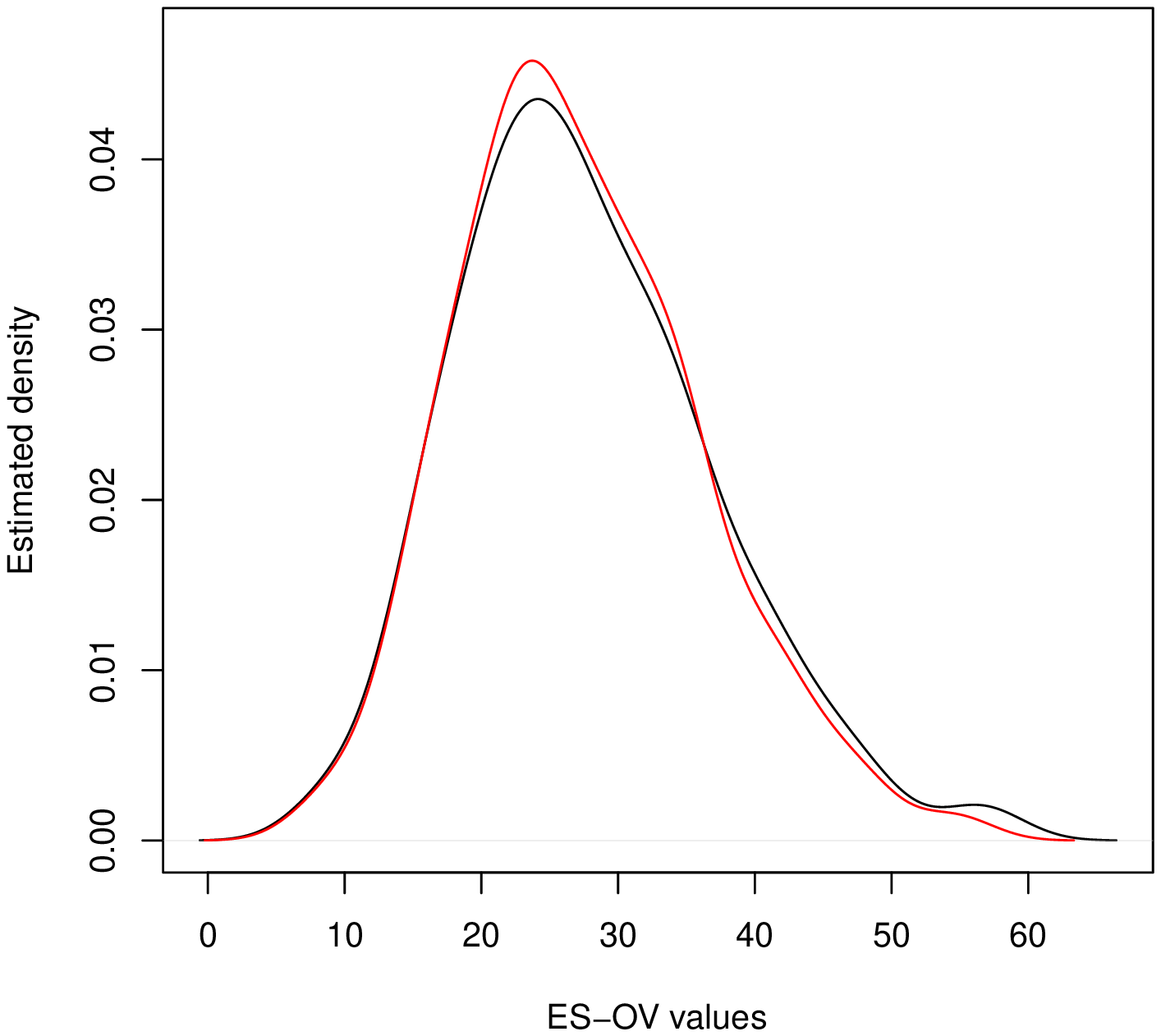} \\
$n=25$ & $n=50$ & $n=75$ & $n=100$                     \\
\multicolumn{4}{c}{16 components} \\ 
\includegraphics[scale=0.25,trim=0 20 0 20]{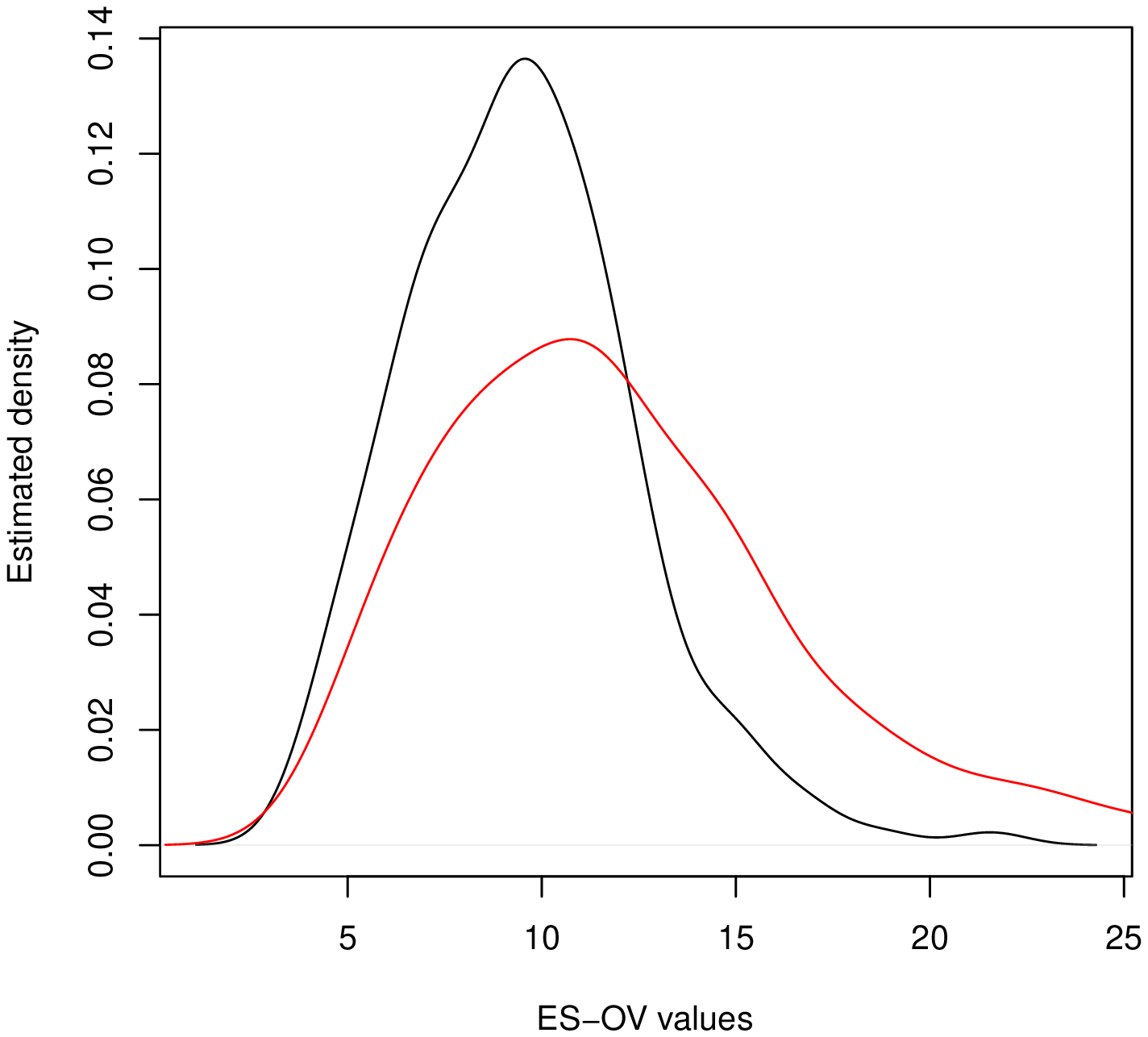} &
\includegraphics[scale=0.25,trim=0 20 0 20]{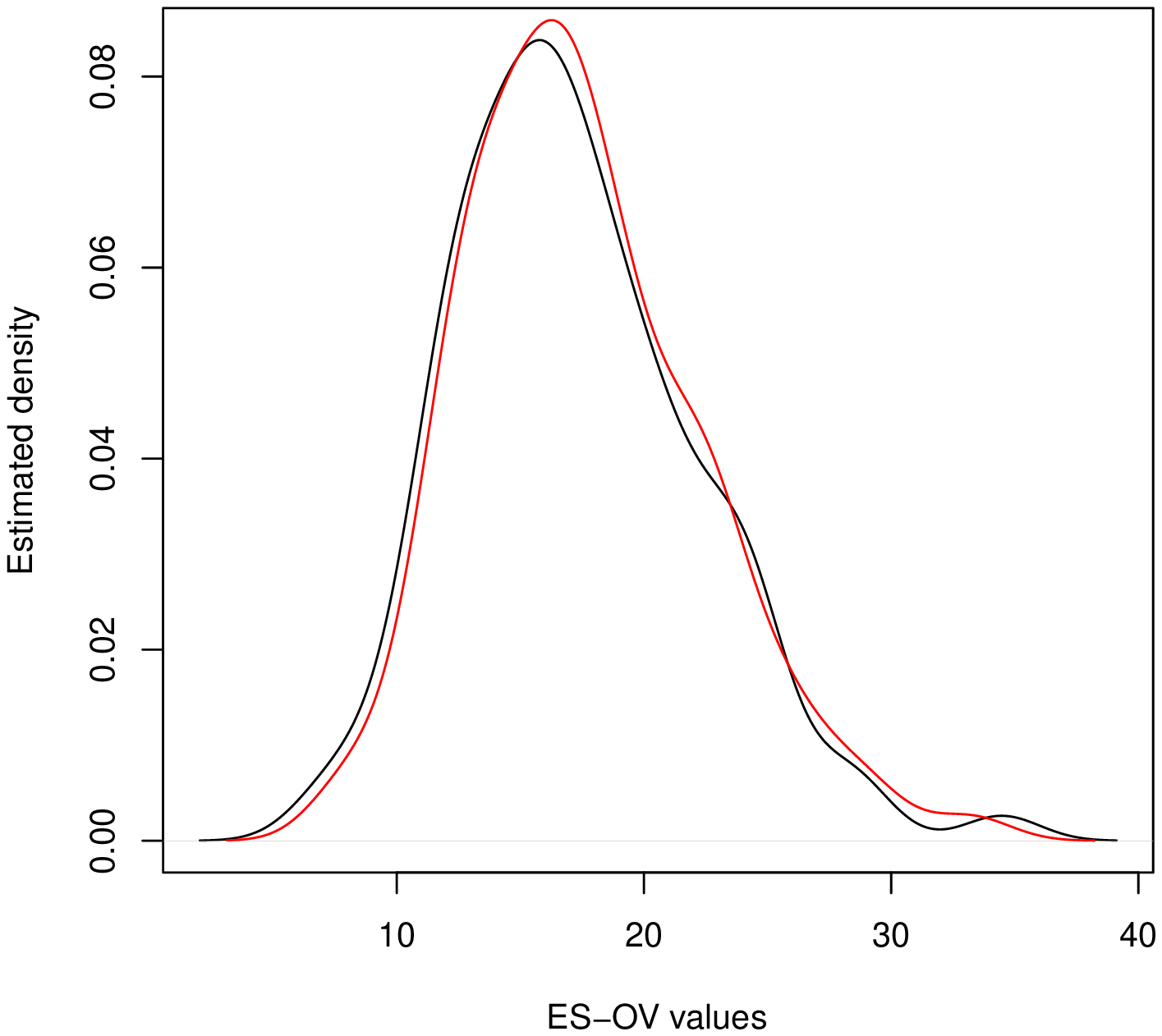} &
\includegraphics[scale=0.25,trim=0 20 0 20]{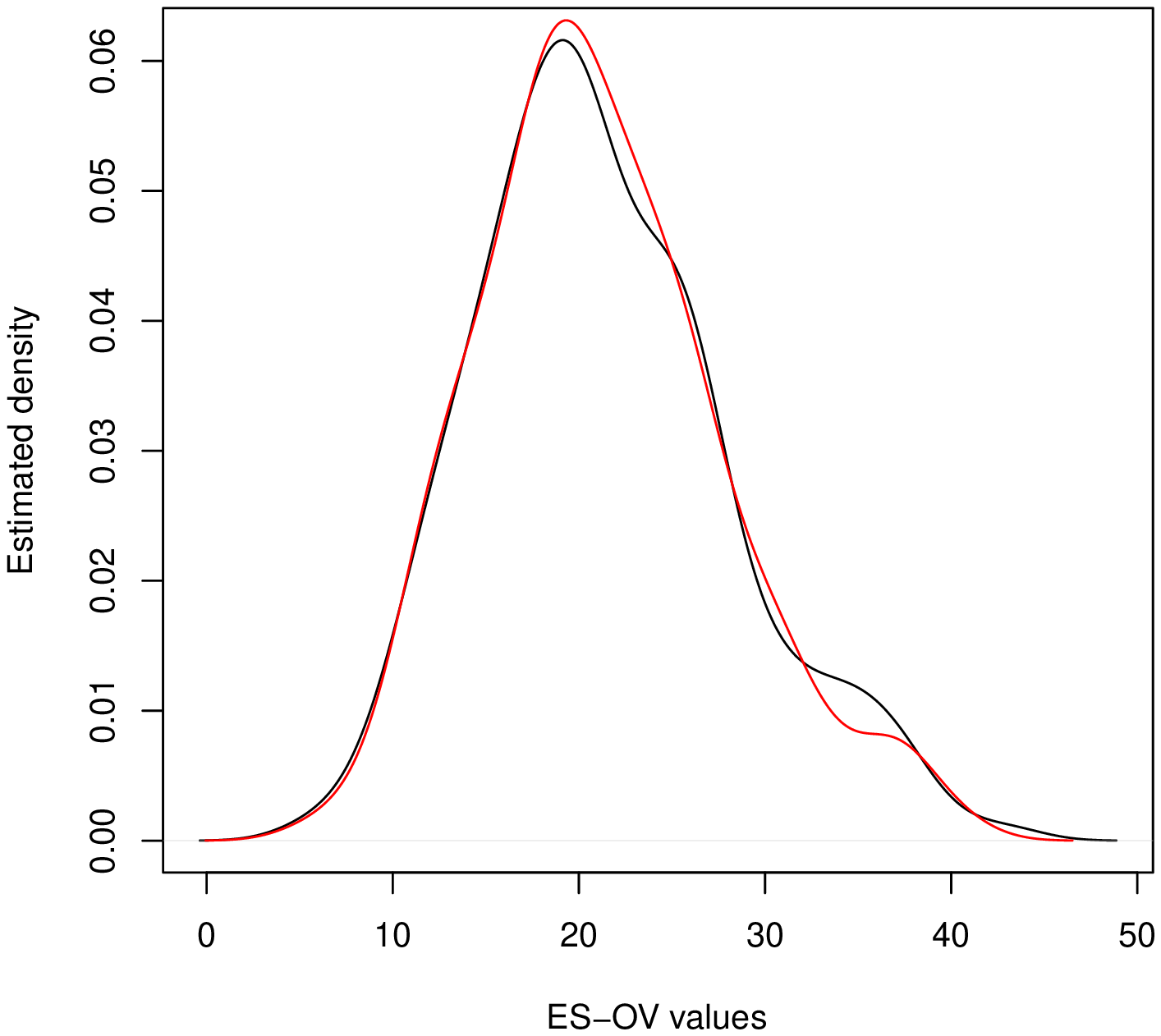} &
\includegraphics[scale=0.25,trim=0 20 0 20]{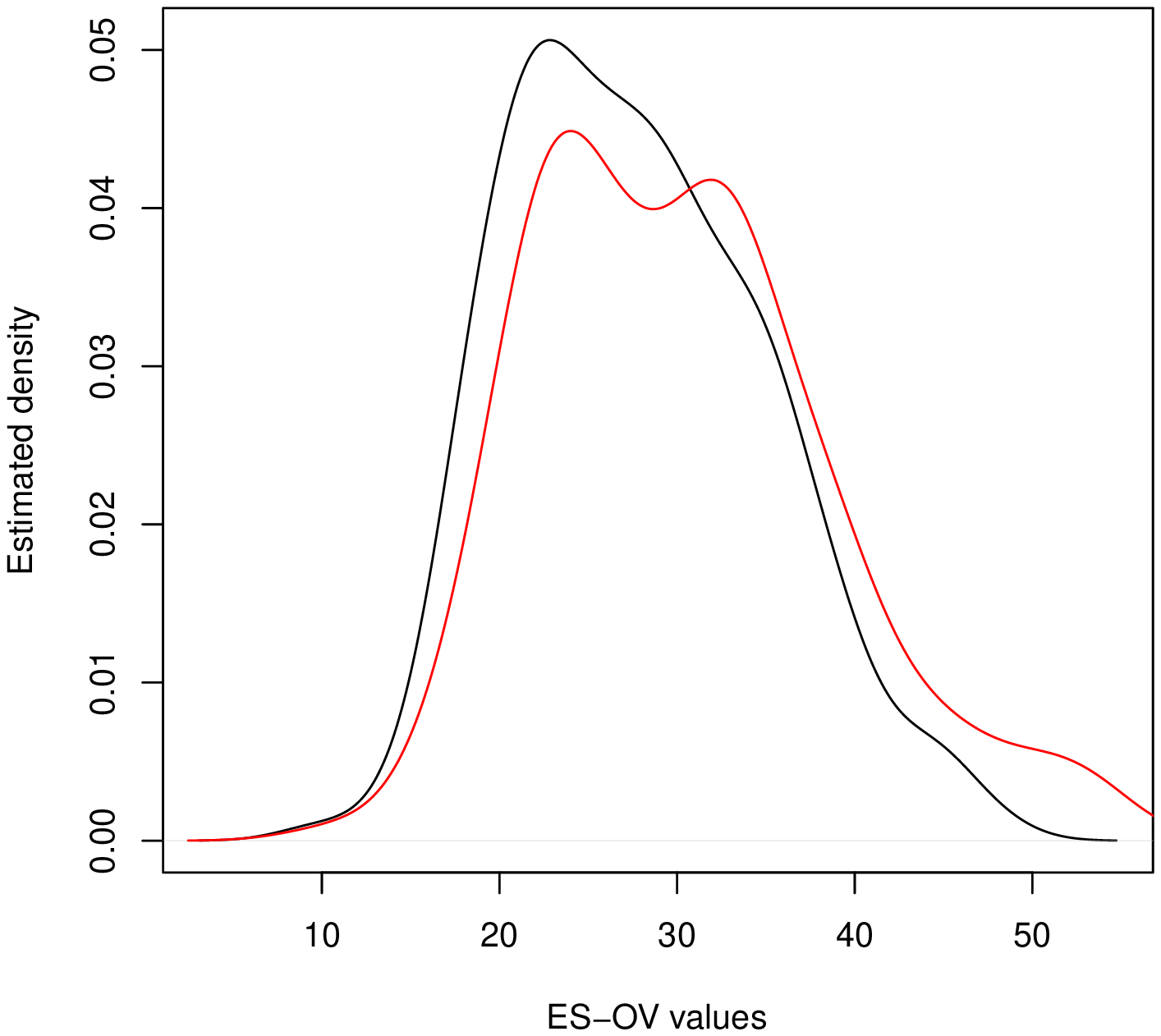}  \\
$n=25$ & $n=50$ & $n=75$ & $n=100$ 
\end{tabular}
\caption{All graphs present the kernel estimated densities of the $200$ Kullback-Leibler divergences of the ES-OV (red line) and the Aitchison regression (black line).}
\label{fig1}
\end{figure}

The results of this simulation study are presented in Table 1. We can see that the proportion of times, the Kulback-Lleibler divergence for the fitted values of the ES-OV regression is smaller than the fitted values of the Aitchison regression, grows large as the sample size increases, but when the sample size is equal to $100$ decays. An explanation that can be given is due to random error. In addition, if we see Figure \ref{fig1} we will see that the Kulback-Leibler divergences of the two regression models are close in general. 

\begin{table}[!ht]
\begin{small}
\begin{center}
\begin{tabular}{c|c|c|c} 
                & \multicolumn{3}{c}{Number of components} \\ \hline
Sample sizes    & 6     & 11    & 16    \\ \hline
$n=25$  & 0.195 & 0.105 & 0.055 \\
$n=50$  & 0.615 & 0.430 & 0.345 \\
$n=75$  & 0.820 & 0.705 & 0.565 \\
$n=100$ & 0.670 & 0.580 & 0.294 \\ \hline
\end{tabular}
\caption{Proportion of times the Kullback-Leibler divergence, of the true values from the fitted values, is smaller for the ES-OV than for the Aitchison regression. The data have no zero values.}
\end{center}
\end{small}
\end{table}

\subsubsection*{Compositional data with zeros}
The calculations are the same as before but the scenario of the simulations has one modification: some values were set to zero
\begin{enumerate}
\item[Step 1.] Generate data from a multivariate normal regression model ${\bf Z} \sim N_p\left({\bf BX},\pmb{\Sigma}\right)$,
where ${\bf X}$ is a design matrix with 2 independent variables. $\bf B$ were chosen randomly from a standard normal distribution and $\pmb{\Sigma}$ was a diagonal matrix with variances generated from an $\text{Exp}(1)$ distribution.
\item[Step 2.] Make the ${\bf Z}$ compositional using
\begin{eqnarray*}
y_1 = \frac{1}{1+\sum_{j=1}^pe^{z_j}}, \ \ y_i = \frac{e^{z_i}}{1+\sum_{j=1}^pe^{z_j}}, \ \ \text{for} \ \ i=2,\ldots,p.
\end{eqnarray*}
\item[Step 3.] Set about $50\%$ (use a uniform distribution) of the elements of these components equal to zero and rescale these vectors so that their sum is again equal to $1$. 
\item[Step 4.] Perform the ES-OV regression and calculate (\ref{akl}).
\item[Step 5.] Use the R package \textit{robCompositions} to replace the zero values and then apply the Aitchison regression and calculate (\ref{akl}). 
\item[Step 6.] Repeat Steps 1-5 for various small sample sizes $n=(25,50,75,100)$ and $p=(5,10,15)$. Thus, we now have $D=(6,11,16)$ number of components and essentially $15, 30$ and $45$ beta parameters to estimate.
\end{enumerate}

Table 2 presents the the proportion of times the Kullback-Leibler divergence for the fitted values of the ES-OV regression is smaller than for the fitted values of the Aitchison regression. The results are clearer now, obviously the ES-Ov has managed to give better predictions, in terms of smaller Kullback-Leibler divergences. 

\begin{table}[!ht]
\begin{small}
\begin{center}
\begin{tabular}{c|c|c|c} 
                & \multicolumn{3}{c}{Number of components} \\ \hline
Sample sizes    & 6     & 11    & 16    \\ \hline
$n=25$  & 0.885 & 0.885 & 0.790  \\
$n=50$  & 0.975 & 0.995 & 0.995  \\
$n=75$  & 1.000 & 1.000 & 1.000  \\
$n=100$ & 1.000 & 1.000 & 0.990  \\ \hline
\end{tabular}
\caption{Proportion of times the Kullback-Leibler divergence, of the true values from the fitted values, is smaller for the ES-OV than for the Aitchison regression. The data have zero values.}
\end{center}
\end{small}
\end{table}

\begin{figure}[!ht]
\centering
\begin{tabular}{cccc}
\multicolumn{4}{c}{6 components} \\ 
\includegraphics[scale=0.25,trim=0 20 0 20]{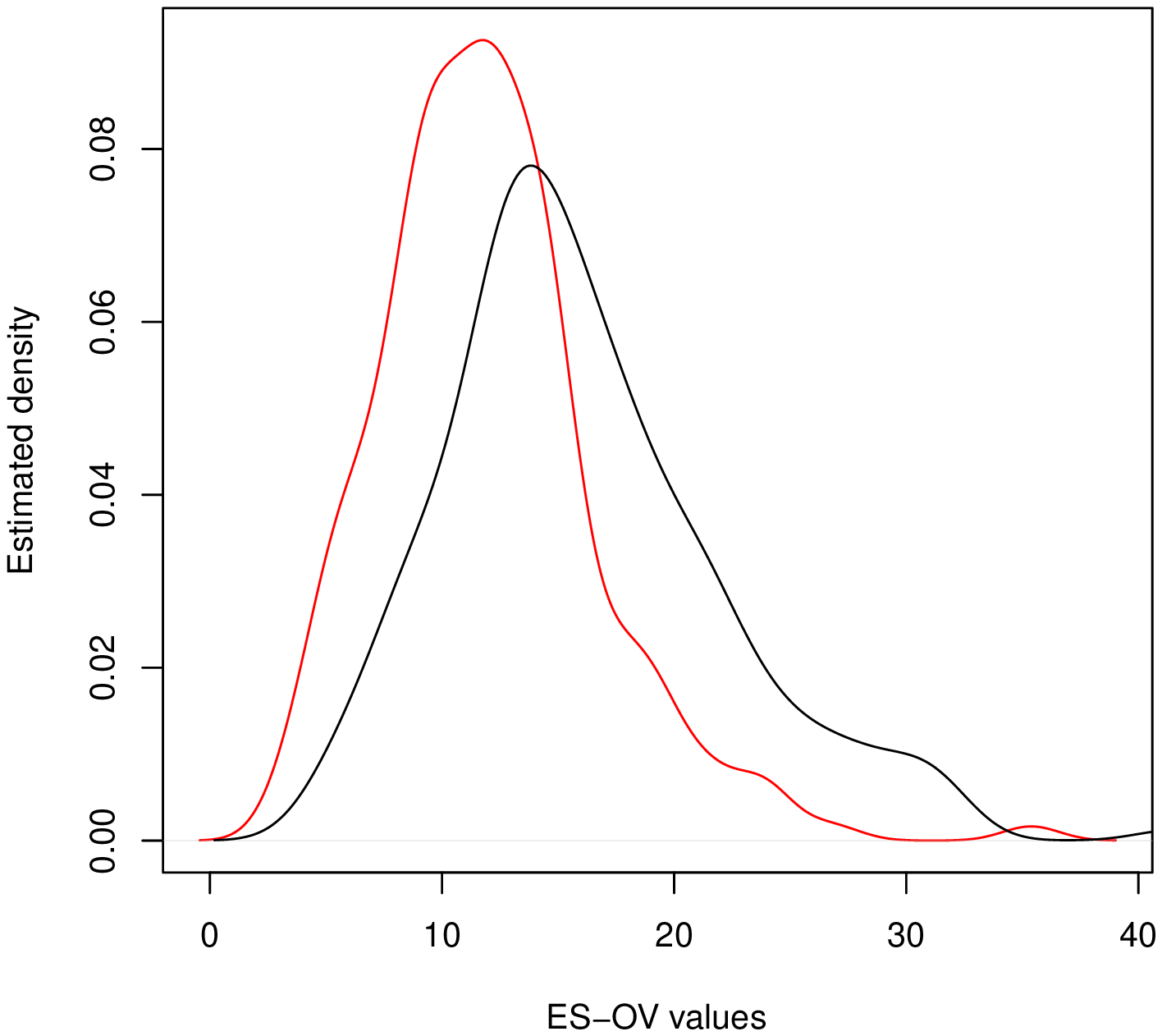} &
\includegraphics[scale=0.25,trim=0 20 0 20]{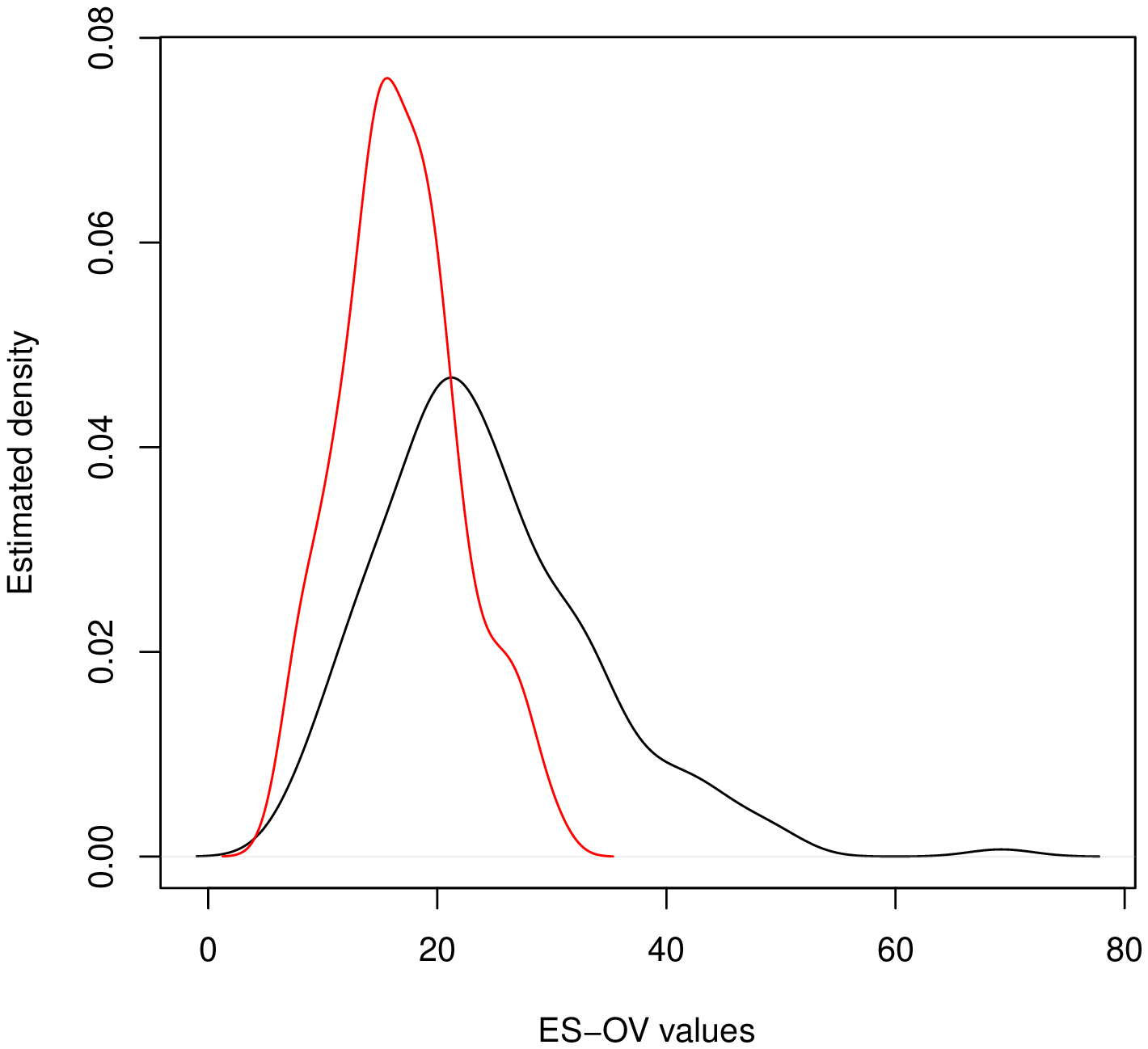} &
\includegraphics[scale=0.25,trim=0 20 0 20]{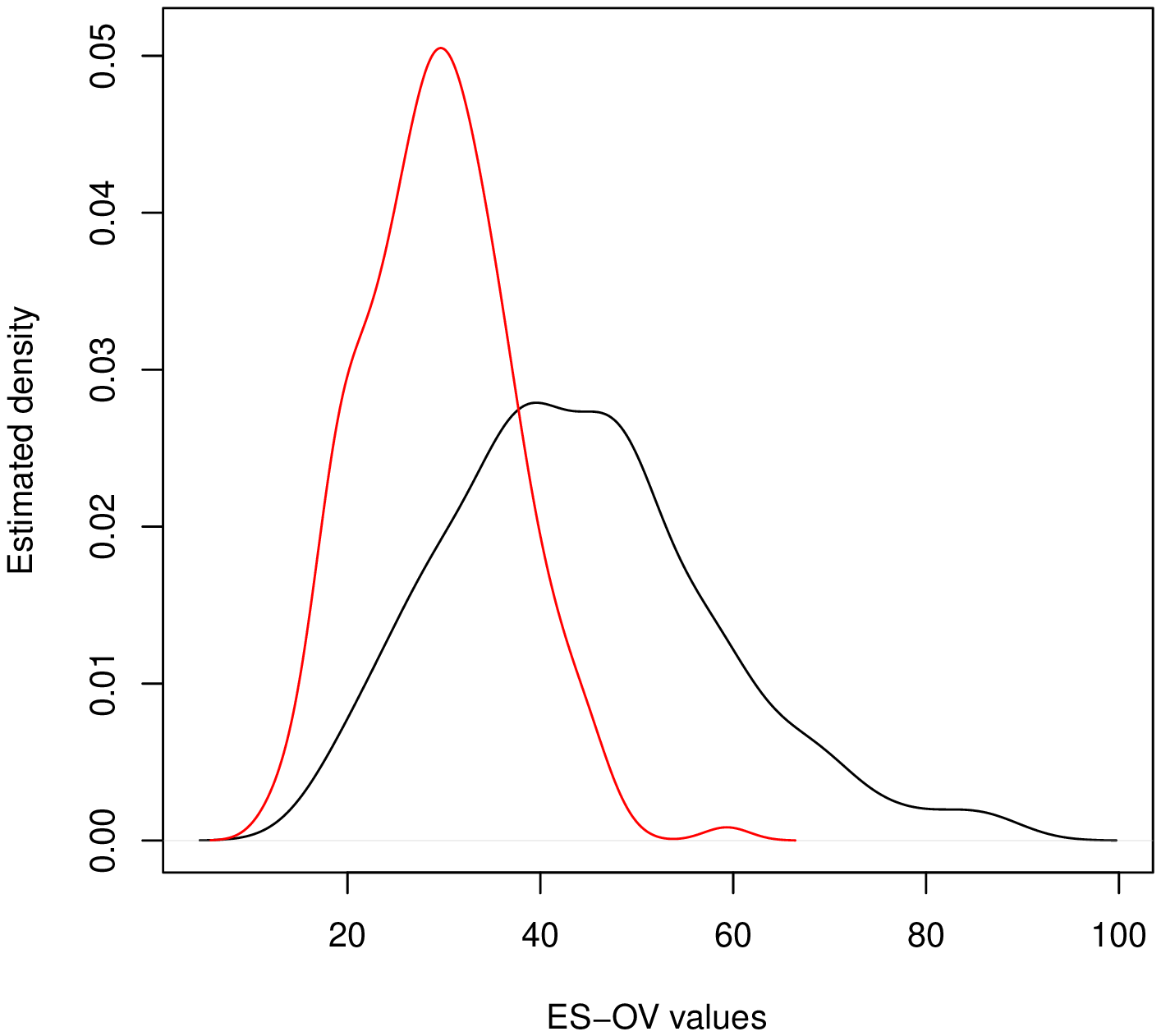} &
\includegraphics[scale=0.25,trim=0 20 0 20]{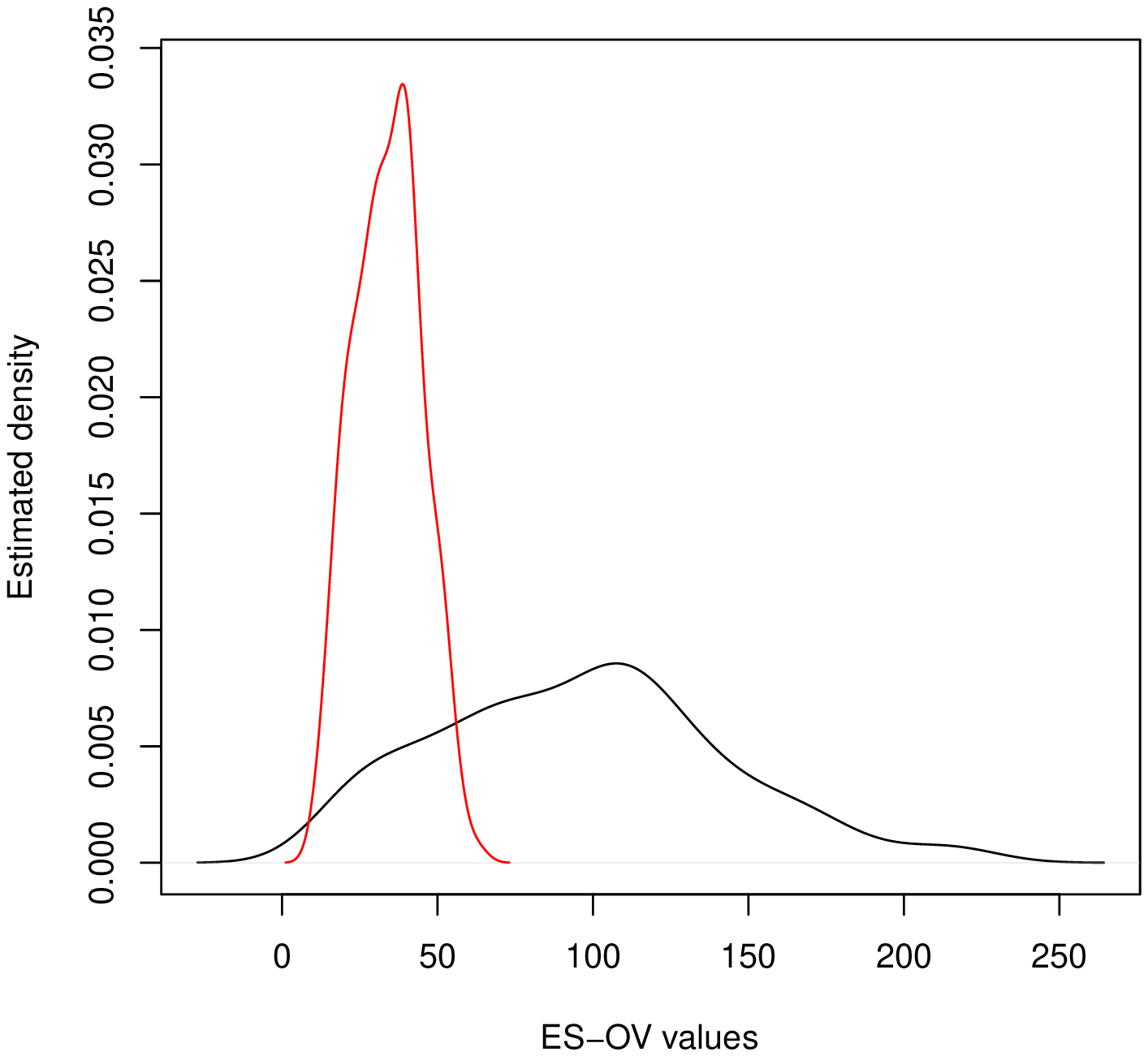} \\
$n=25$ & $n=50$ & $n=75$ & $n=100$                     \\
\multicolumn{4}{c}{11 components} \\ 
\includegraphics[scale=0.25,trim=0 20 0 20]{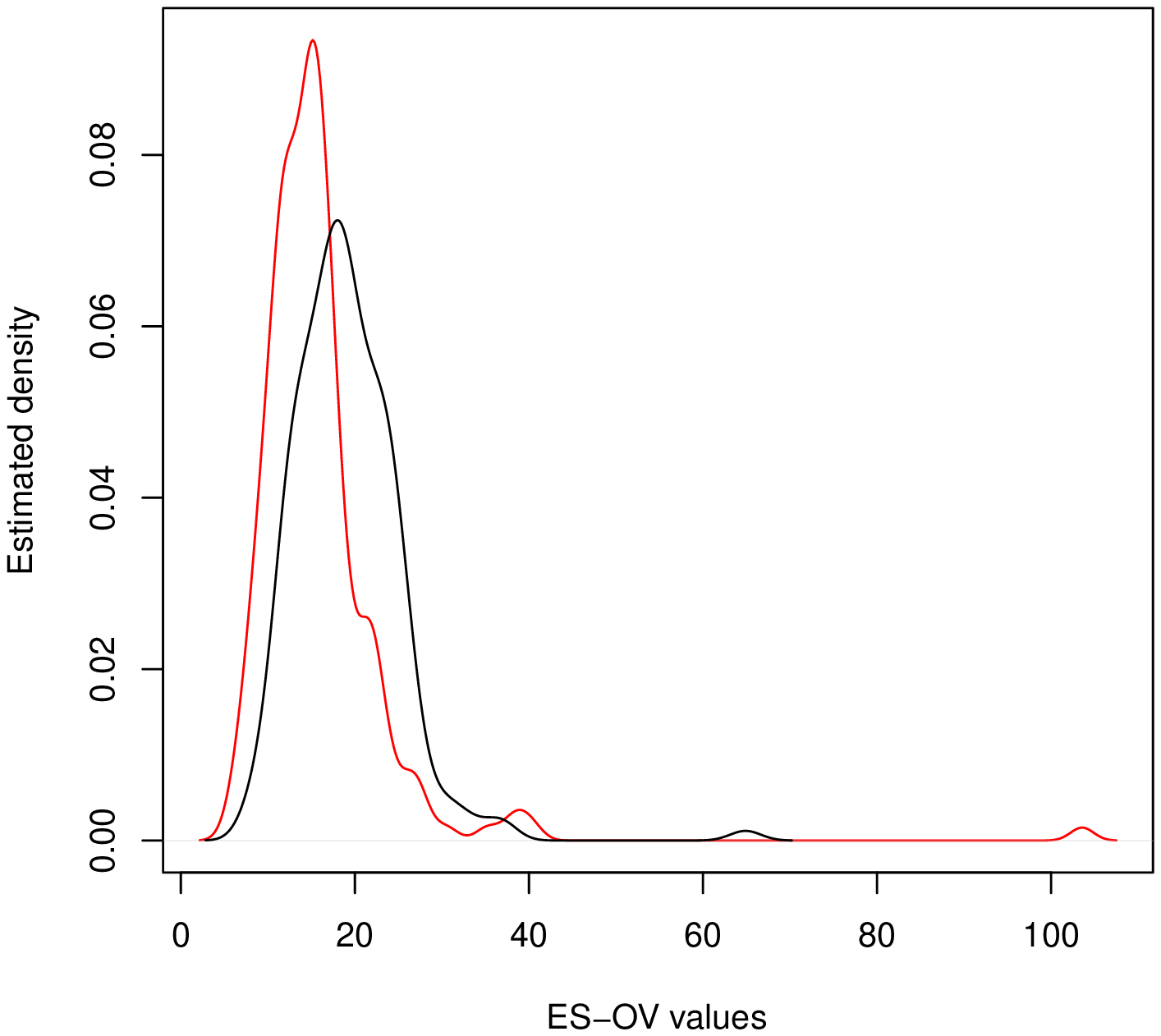} &
\includegraphics[scale=0.25,trim=0 20 0 20]{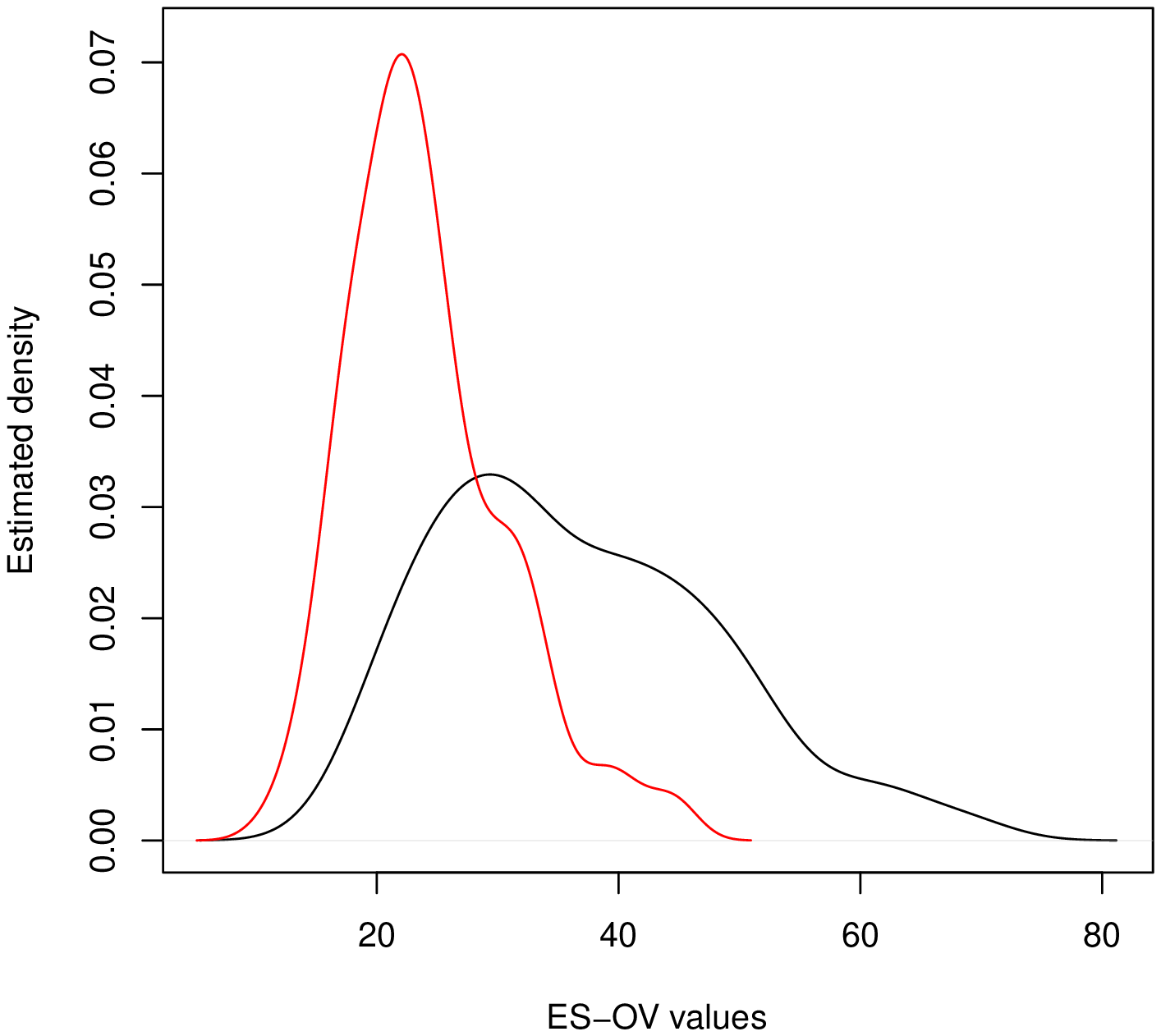} &
\includegraphics[scale=0.25,trim=0 20 0 20]{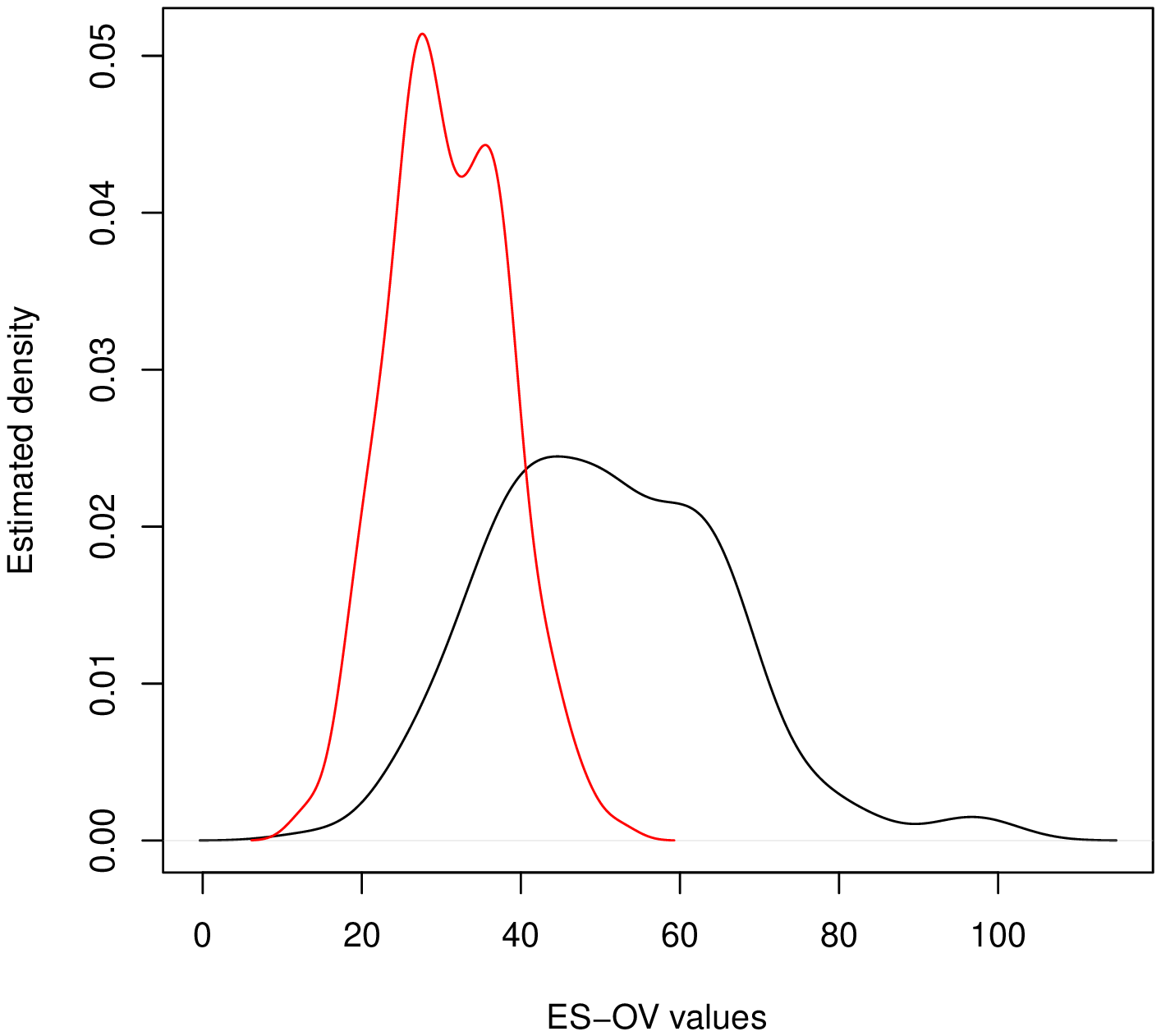} &
\includegraphics[scale=0.25,trim=0 20 0 20]{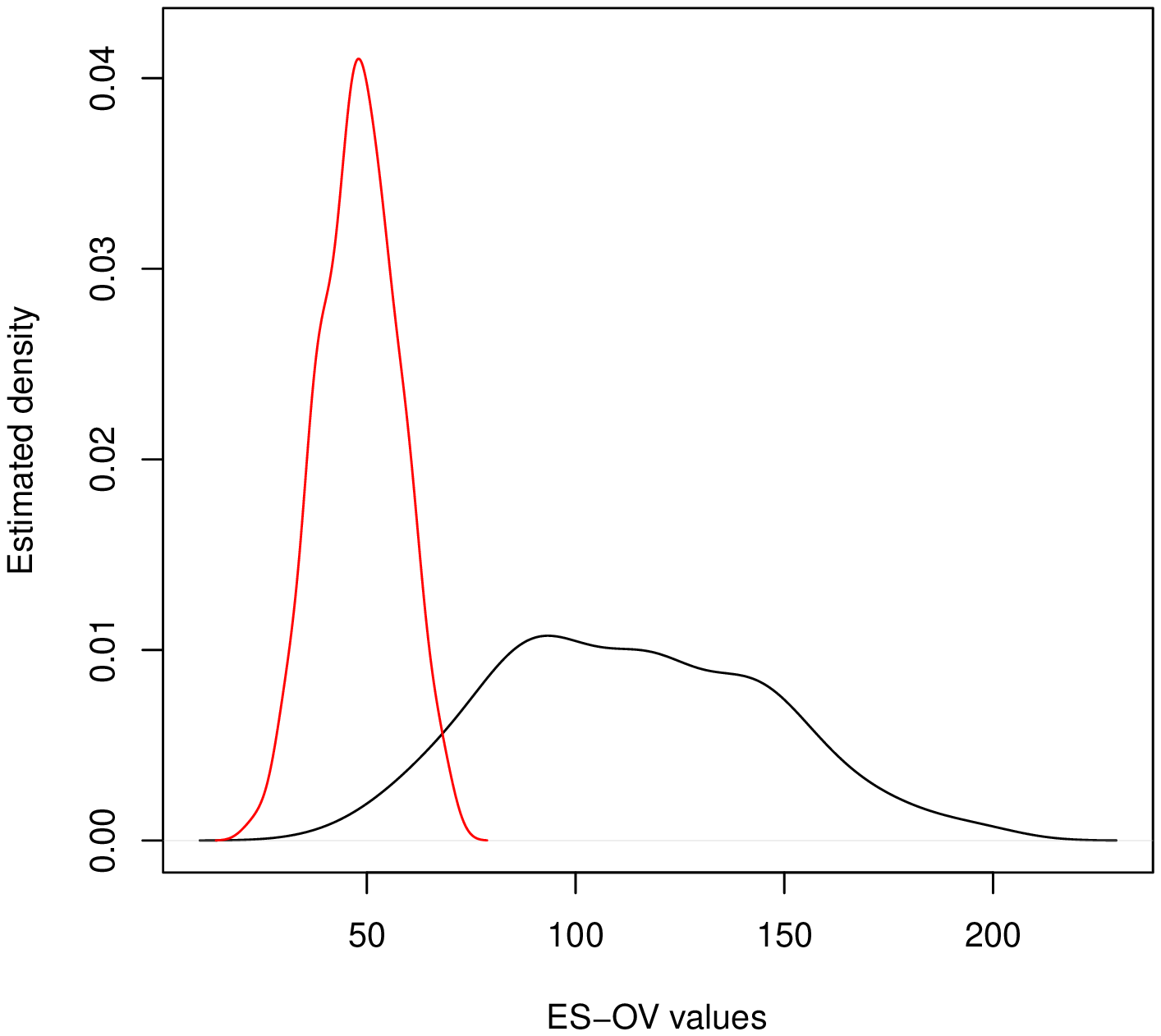} \\
$n=25$ & $n=50$ & $n=75$ & $n=100$                     \\
\multicolumn{4}{c}{16 components} \\ 
\includegraphics[scale=0.25,trim=0 20 0 20]{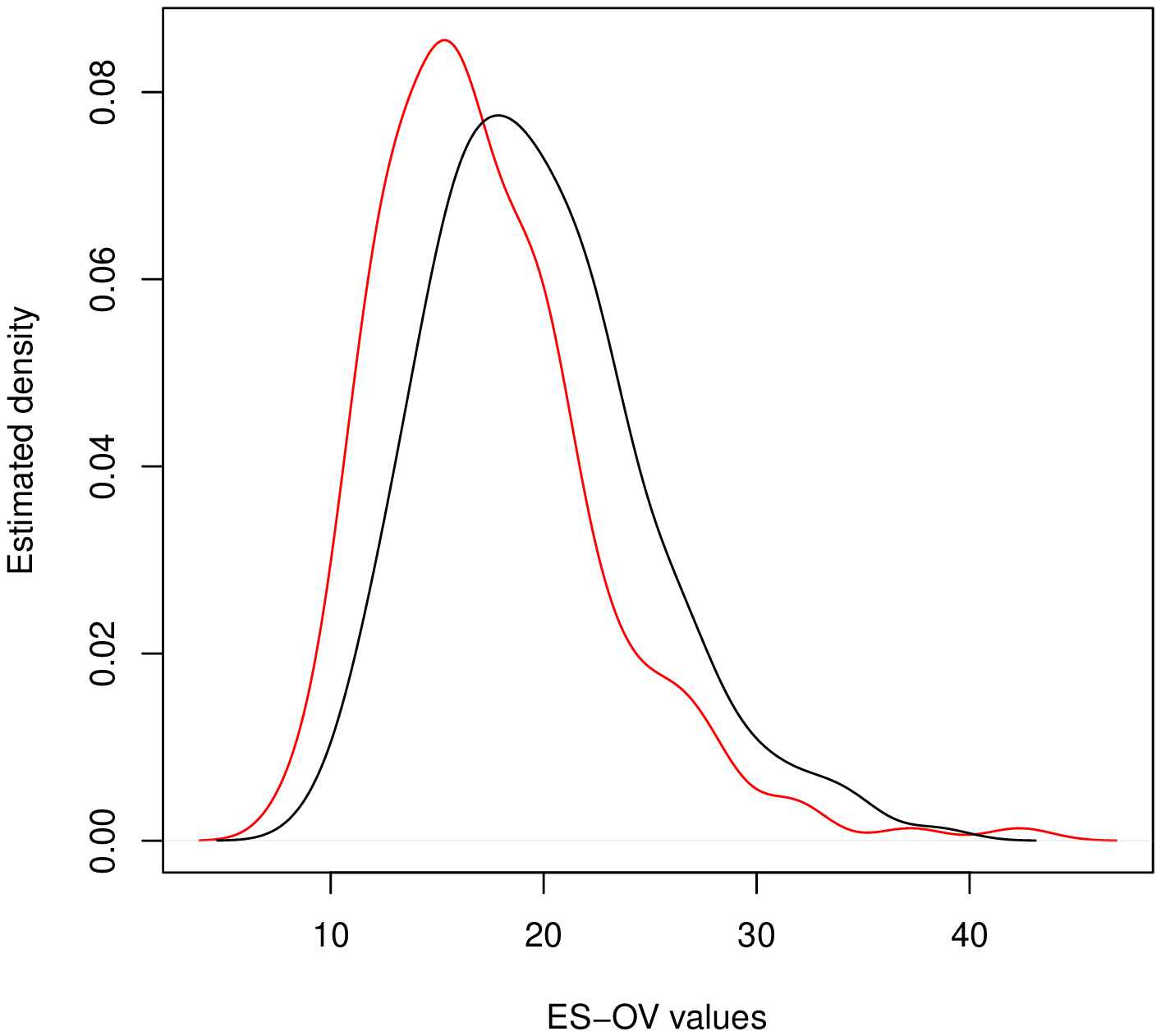} &
\includegraphics[scale=0.25,trim=0 20 0 20]{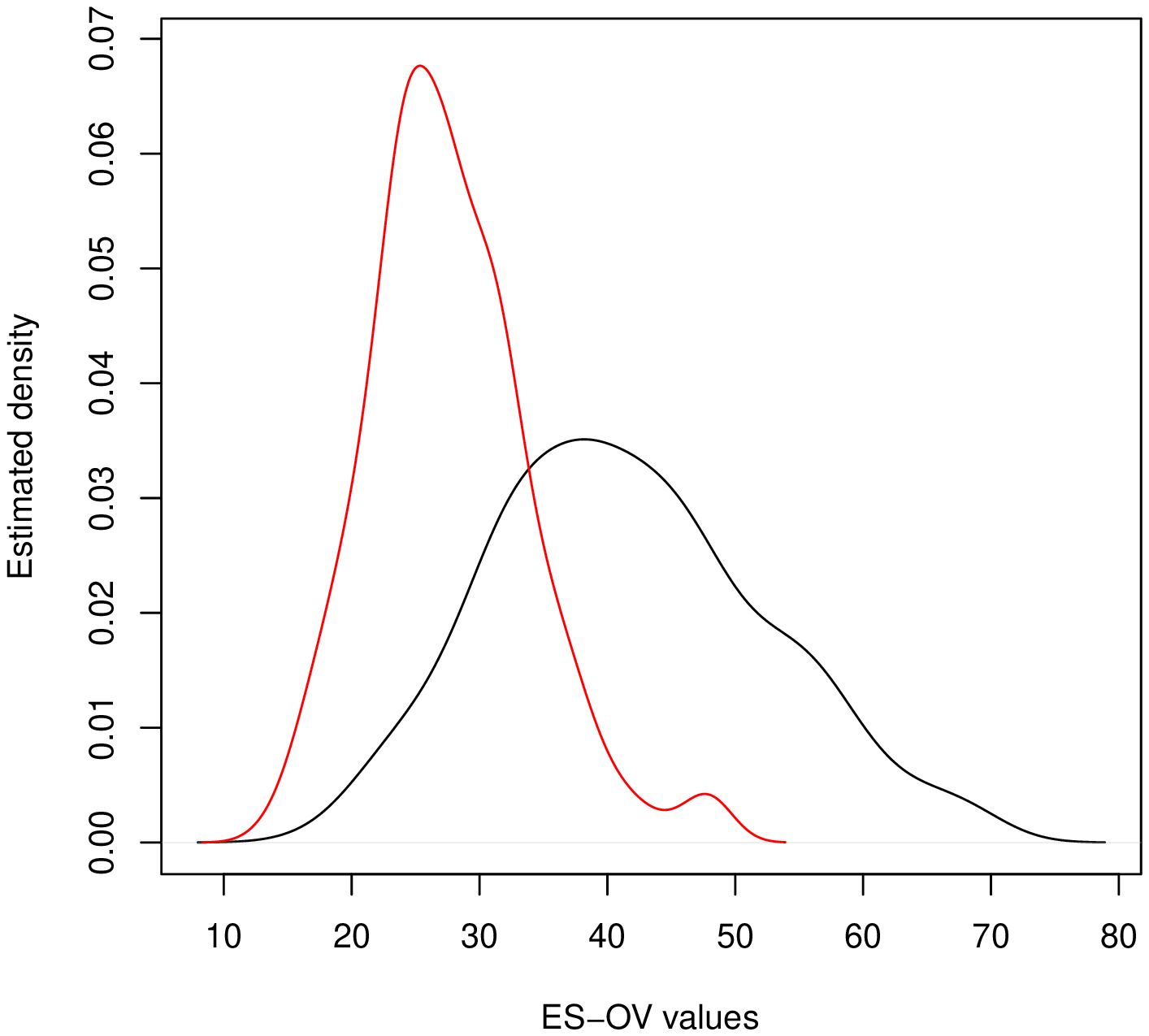} &
\includegraphics[scale=0.25,trim=0 20 0 20]{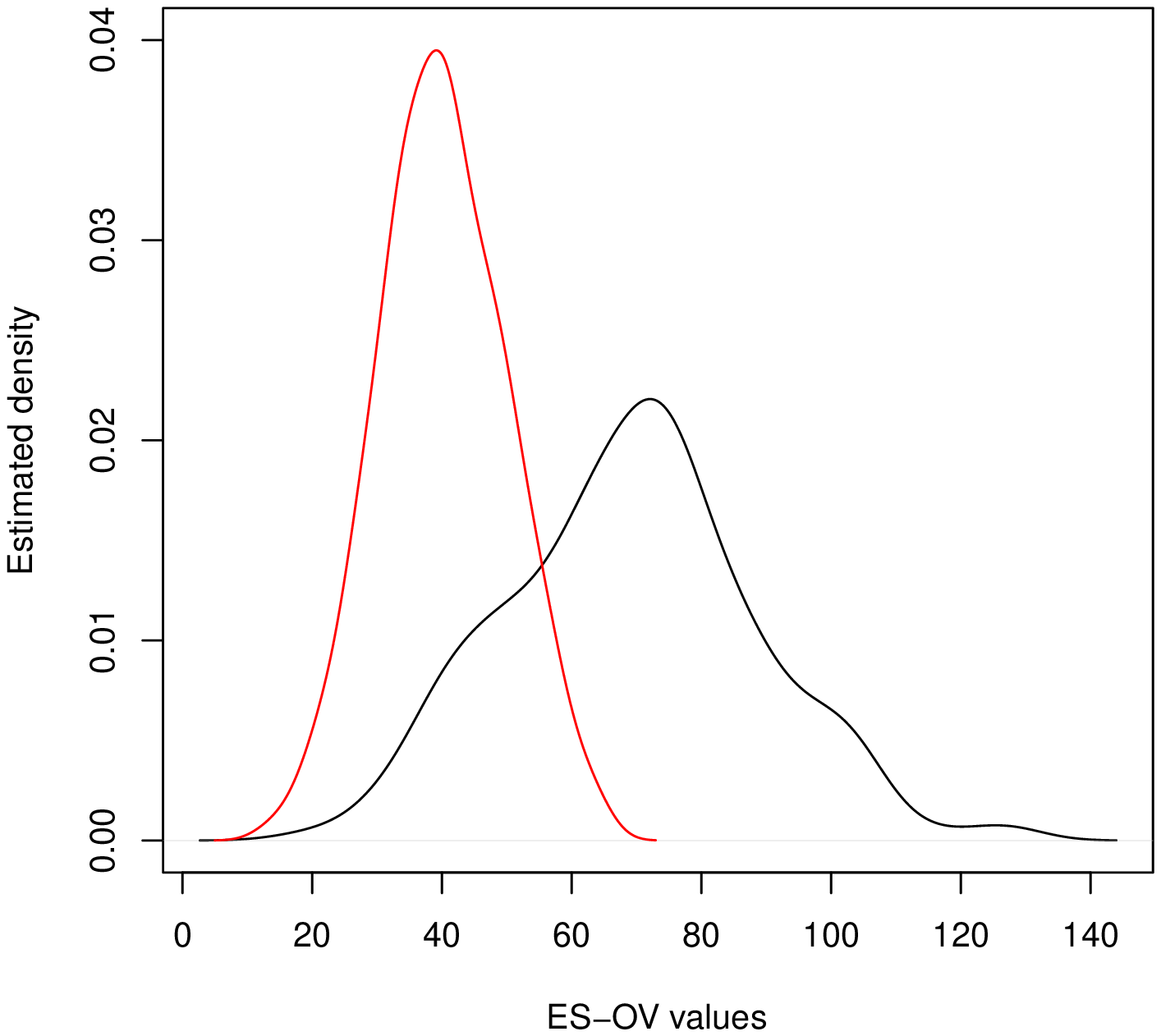} &
\includegraphics[scale=0.25,trim=0 20 0 20]{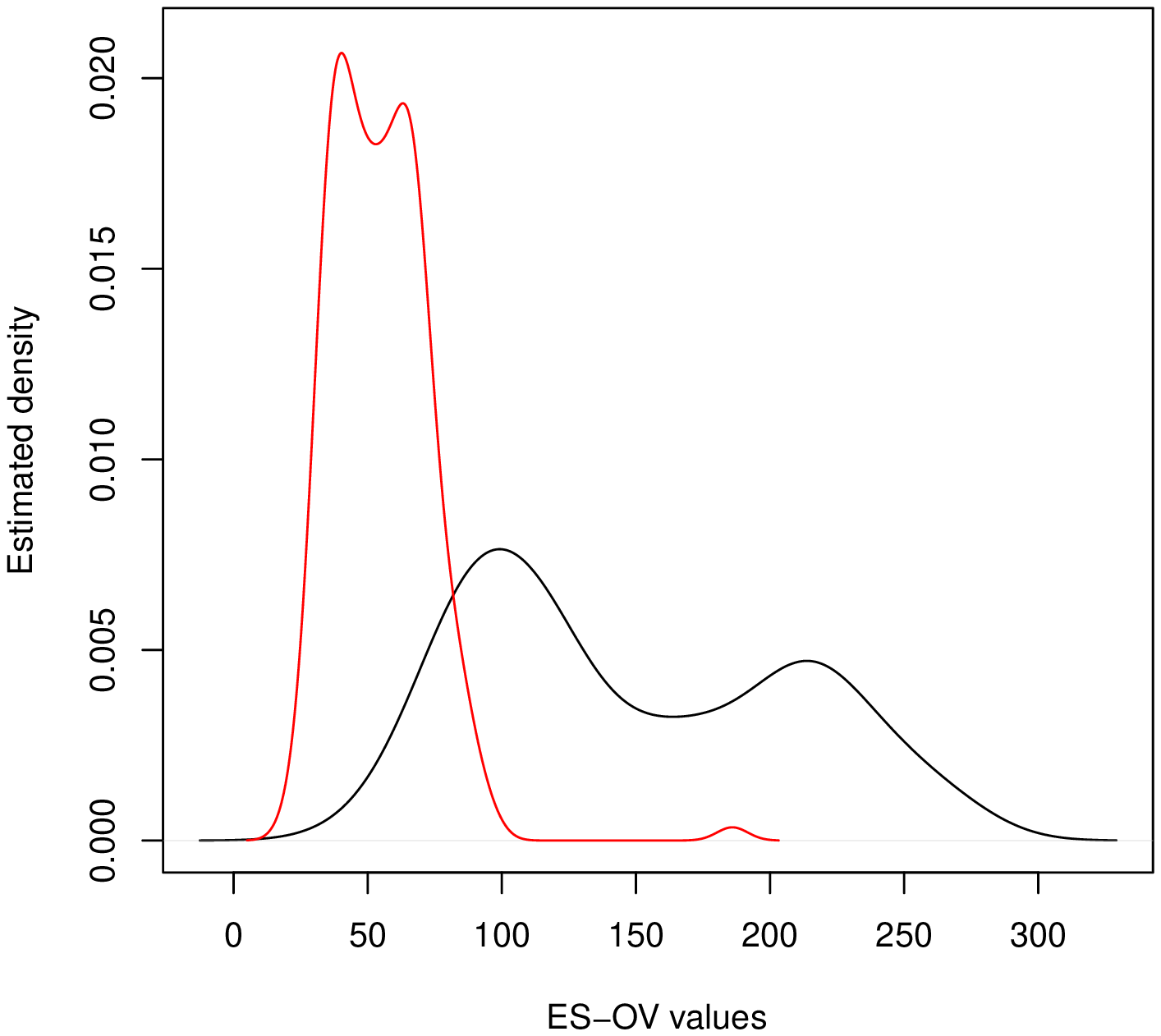}  \\
$n=25$ & $n=50$ & $n=75$ & $n=100$ 
\end{tabular}
\caption{All graphs present the kernel estimated densities of the $200$ Kullback-Leibler divergences of the ES-OV (red line) and the Aitchison regression (black line).}
\label{fig2}
\end{figure}

\section{Miscellanea}
When $D=2$, we end up with proportional data which can be analysed using a beta distribution. Kateri and Agresti (2010) used a $\phi$-divergence regression model for binary responses. We can say, that in this cases the ES-OV is a special case of that model, since the ES-OV belongs to the class of $\phi$-divergence statistics as we saw before. 

Can we use another $\phi$-divergence statistic for regression. Of course we can, the $\chi^2$ distance and the Hellinger distance have been used and the results were very good. Again, the issue of consistency needs to be checked. Until this is done, they can be used for prediction purposes. In both of these cases, statistical properties about the parameters in the discrete case has been established. What about the compositional data case? 
 
What happens when compositional data exist in the independent variables side? We request only for one condition, no zeros to exist in the independent variables compositional data. If this is satisfied, one can apply the centred log-ratio transformation (Atchison, 2003) to the independent side compositional data
\begin{eqnarray*}
z_i=\log{\frac{x_i}{g\left({\bf x}\right)}}, \ \ \text{for} \ \ i=1,\ldots,D,
\end{eqnarray*}
where $g\left({\bf x}\right)=\prod_{i=1}^{1/D}$ is the geometric mean of the components. Then, multiply the transformed data with the Helmert sub-matrix (Lee and Small, 1999), which is an orthonormal $D\times D$ matrix (Lancaster, 1965), whose first row, proportional to the vector of $1s$, is deleted. This removes the singularity problem. Half of the job is done. Now, one can perform PCA on ${\bf Z}^T{\bf Z}$ and use the scores of the principal components, which is essentially a principal component regression (Jolliffe 2005). The number of principal components to be used can be tuned via cross validation by minimizing the Kullback-Leibler divergence. 

\section{Conclusions}
We cannot say that we have proposed a new regression model that always outperforms the Aitchison regression. In fact we did not even see how it compares to the other methods mentioned in Section 2. We saw an example though where Aitchison regression fails. We also saw that in the case of many zeros, Aitchison regression is not very robust, but we believe this is a problem of the EM algorithm for the zero value replacements (Templ 2011) and not of the regression itself. For these cases, the ES-OV regression does better than Aitchison regression. The goal was not to find a model that outperforms always all the other existing ones, but to propose one which is equally good or better in some cases. In the case of no zero values, Aitchison regression seems to do a better job, but in the case of many zeros, ES-OV regression is suggested. 

In addition, since we have not examined the asymptotic properties of the ES-OV regression, this method is currently advisable to be used for prediction purposes. So, if the interest is to estimate as accurately as possible the compositions of new observations, given a set of observed data containing α lot of zeros, we suggest the use of the ES-OV regression. 
  
\subsubsection*{Acknowledgements}
The author would like to express his gratitude to Aziz Alenazi (PhD student in statistics at the university of Sheffield) and to Theo Kypraios (Assistant professor in statistics at the university of Nottingham) for their help with the computations.

\section*{Appendix: R codes}

\begin{verbatim}

### Aitchison regression

compreg <- function(y, x) {
  ## y is dependent variable, the compositional data
  ## x is the independent variable(s)
  ## the next two commands make sure the data are matrices 
  y <- as.matrix(y)   ## makes sure y is a matrix 
  y <- y/rowSums(y)  ## makes sure y is compositional data 
  z <- log( y[, -1] / y[, 1] )  ## additive log-ratio (alr) transformation 
  ## with the first component being the base
  n <- nrow(z)  ## sample size
  d <- ncol(z)  ## dimensionality of z
  p <- ncol(x)  ## dimensionality of x
  X <- as.matrix( cbind(1, x) )  ## the design matrix
  beta <- solve(t(X) %*% X) %*% t(X) %*% z  ## the parameters
  est1 <- X %*% beta ## fitted values
  est2 <- cbind(1, exp(est1))
  est <- est2 / rowSums(est2) 
  list(beta = beta, fitted = est) 
}


### ES-OV regression

esov.compreg <- function(y, x) {
  ## y is dependent variable, the compositional data
  ## x is the independent variable(s) 
  y <- as.matrix(y) 
  y <- y/rowSums(y)  ## makes sure y is compositional data 
  x <- as.matrix( cbind(1, x) )
  d <- ncol(y) - 1  ## dimensionality of the simplex
  n <- nrow(y)  ## sample size
  z <- list(y = y, x = x)
  reg <- function(para, z = z){
    y <- z$y  ;  x <- z$x
    be <- matrix(para,byrow = T,ncol = d)
    mu1 <- cbind(1, exp(x %*% be))
    mu <- mu1/rowSums(mu1)
    M <- (mu + y)/2
    f <- sum( y * log(y / M) + mu * log(mu / M), na.rm = T )
    f 
  }
  ## the next lines minimize the reg function and obtain the estimated betas
  ini <- as.vector( t( coef( lm(y[, -1] ~ x[, -1]) ) ) )  ## initial values
  val <- NULL
  qa <- nlm(reg, ini, z = z) 
  val[1] <- qa$minimum
  qa <- nlm(reg, qa$estimate, z = z)
  val[2] <- qa$minimum
  i <- 2
  while (val[i-1] - val[i] > 0.00001) {
    i <- i + 1
    qa <- nlm(reg, qa$estimate, z = z)
    val[i] <- qa$minimum 
  }
  val <- min(val)
  beta <- matrix(qa$estimate, byrow = T, ncol = d)
  mu1 <- cbind(1, exp(x %*% beta))
  mu <- mu1 / rowSums(mu1)
  list(beta = beta, val = val, fitted = mu)
}
\end{verbatim}

\vspace{5pt}
\begin{center}
{\large {\bf REFERENCES} }
\end{center}

\vspace*{4pt}

\begin{description}

\item{ Aitchison, J. (1982). The statistical analysis of compositional data. \it{Journal of the Royal
Statistical Society. Series B} {\bf 44}, 139-177.}

\item { Aitchison, J. (2003). \it{ The statistical analysis of compositional data,} New Jersey: Reprinted by The Blackburn
Press. }

\item{ Endres, D. M. and Schindelin, J. E. (2003). A new metric for probability distributions. \it{ Information Theory, IEEE Transactions on}  {\bf 49}, 1858-1860.}

\item{ Gueorguieva, R., Rosenheck, R., and Zelterman, D. (2008). Dirichlet component regression and
its applications to psychiatric data. \it{ Computational statistics \& data analysis} {\bf 52}, 5344-5355.}

\item{ Jolliffe, I. T. (2005). \it{Principal component analysis, } New York: Springer-Verlag.}

\item{ Kateri, M. and Agresti, A. (2010). A generalized regression model for a binary response.
\it{Statistics \& Probability Letters} {\bf 80}, 89-95.}

\item{ Kent, J. T. (1982). The Fisher-Bingham distribution on the sphere. \it{ Journal of the Royal
Statistical Society. Series B} {\bf 44}, 71-80}.

\item{ Kullback, S. (1997). \it{Information theory and statistics}, New York: Dover Publications.}

\item{ Lancaster, H. (1965). The Helmert matrices. \it{American Mathematical Monthly} {\bf 72}, 4-12.}

\item{ Le, H. and Small, C. G. (1999). Multidimensional scaling of simplex shapes. \it{Pattern Recognition}
{\bf 32}, 1601-1613.}

\item{ Maier, M. J. (2014). DirichletReg: Dirichlet Regression in R. \it{R package version 0.5-2} .}

\item{ Mart{\'\i}n-Fern{\'a}ndez, J., Hron, K., Templ, M., Filzmoser, P., and Palarea-Albaladejo, J. (2012).
Model-based replacement of rounded zeros in compositional data: Classical and robust approaches. \it{Computational Statistics \& Data Analysis} {\bf 56}, 2688-2704.}

\item{ Murteira, J. M. and Ramalho, J. J. (2014). Regression analysis of multivariate fractional data.
\it{Econometric Reviews} (ahead-of-print) 1-38.}

\item{ Neocleous, T., Aitken, C., and Zadora, G. (2011). Transformations for compositional data
with zeros with an application to forensic evidence evaluation. \it{Chemometrics and IntelligentLaboratory Systems} {\bf 109}, 77-85.}

\item{ {\"O}sterreicher, F. and Vajda, I. (2003). A new class of metric divergences on probability spaces
and its applicability in statistics. \it{Annals of the Institute of Statistical Mathematics} {\bf 55}, 639-653.}

\item{ Scealy, J. L. and Welsh, A. H. (2011). Regression for compositional data by using distributions
defined on the hypersphere. \it{Journal of the Royal Statistical Society. Series B} {\bf 73}, 351-375.}

\item{ Stephens, M. A. (1982). Use of the von Mises distribution to analyse continuous proportions.
\it{Biometrika} {\bf 69}, 197-203.}

\item{ Templ, M., Hron, K., and Filzmoser, P. (2011). robCompositions: Robust estimation for compositional data. \it{R package version 0.8-4}.}

\item{ Theil, H. (1967). \it{Economics and information theory}. Amsterdam: North-Holland publishing
company.}

\end{description}

\end{document}